\documentclass[acmsmall,screen,nonacm]{acmart}
\setcopyright{none}
\usepackage{xcolor}
\usepackage{amsmath}
\usepackage{amsfonts}
\usepackage{booktabs}
\usepackage{array}
\usepackage{longtable}
\usepackage{pdflscape}
\usepackage{enumitem}
\usepackage{graphicx}
\usepackage{subcaption}
\usepackage{tikz}
\usetikzlibrary{arrows.meta,calc}

\title{%
Simulation-Based Testing of RCM-Constrained Robotic Software via Task Manifolds
}

\author{Zhenyu Jiang}
\orcid{0009-0008-2020-586X}
\affiliation{%
  \institution{School of Intelligent Science and Technology, Hangzhou Institute for Advanced Study, University of Chinese Academy of Sciences}
  \city{Hangzhou Shi}
  \state{Zhejiang Sheng}
  \country{China}}
\email{jiangzhenyu25@mails.ucas.ac.cn}

\author{Yanlin Li}
\affiliation{%
  \institution{School of Computer Science and Engineering, Sun Yat-sen University}
  \city{Guangzhou Shi}
  \state{Guangdong Sheng}
  \country{China}}
\email{liylin65@mail2.sysu.edu.cn}

\author{Rongjie Yan}
\correspondingauthor
\affiliation{%
  \institution{Institute of Software Chinese Academy of Sciences}
  \city{Haidian Qu}
  \state{Beijing Shi}
  \country{China}}
\email{yrj@ios.ac.cn}

\author{Rihui Song}
\correspondingauthor
\affiliation{%
  \institution{School of Computer Science and Engineering, Sun Yat-sen University}
  \city{Guangzhou Shi}
  \state{Guangdong Sheng}
  \country{China}}
\email{songrh3@mail.sysu.edu.cn}

\author{Kai Huang}
\correspondingauthor
\affiliation{%
  \institution{School of Computer Science and Engineering, Sun Yat-sen University}
  \city{Guangzhou Shi}
  \state{Guangdong Sheng}
  \country{China}}
\email{huangk36@mail.sysu.edu.cn}

\date{\today}

\begin{document}

\renewcommand{\sectionautorefname}{Section}
\renewcommand{\subsectionautorefname}{Section}
\renewcommand{\subsubsectionautorefname}{Section}

\begin{abstract}
System-level testing of surgical robots requires generating executable test scenarios that exercise diverse, safety-critical behaviors. However, remote-center-of-motion (RCM) surgical manipulators operate within tightly constrained workspaces governed by coupled task geometry, inverse kinematics, and motor limits. Direct sampling in motor-command or Cartesian spaces frequently wastes testing budgets on physically unfeasible configurations. To address this limitation, we present a manifold-centered testing framework that decouples feasible test-domain construction from downstream test-generation strategies. Candidate inputs are generated directly within this constructed feasible domain and mapped to lower-level joint commands for simulation execution. Offline trajectory safety oracles and a spatial reference grid then evaluate failure yield and coverage.
We compare our task-manifold representation against Cartesian XYZ and principal component analysis (PCA) under random testing, systematic traversal, and genetic search strategies.
Results demonstrate that the feasible task domain improves test-generation efficiency and enables systematic reference-cell coverage.
Crucially, random and genetic search strategies detect a broader spectrum of failure patterns than baseline representations, revealing that complete reference-cell coverage alone does not exhaust intra-cell behavioral dynamics.
Furthermore, differential testing between defective and patched controller versions reveals that targeted evolutionary search detects aggregate failure-rate shifts from software repairs. Finally, a cross-platform study on a 7-DoF Kinova Gen3 arm in MoveIt~2 confirms workflow transferability to kinematically redundant manipulators. Together, these findings highlight the necessity of combining feasibility-aware domain modeling with complementary search strategies for systematic testing of robot software.
\end{abstract}



\maketitle

\section{Introduction}

Surgical robots support precise manipulation in confined workspaces
\cite{taylor2003medical,chatterjee_advancements_2024}. For example, in ophthalmic
microsurgery, instruments maneuver through a narrow entry port while avoiding
damage to delicate intraocular tissue
\cite{thirunavukarasu_robot-assisted_2024}. Recent autonomous platforms have demonstrated closed-loop needle navigation and targeted retinal injections~\cite{zhang_autonomous_2023, bian_autonomous_2026}. Executing these delicate interventions reliably demands tightly coupled perception, trajectory planning, and low-level control to reach surgical targets while strictly adhering to kinematic and safety constraints.

Comprehensive system-level testing is essential to ensure the operational safety of surgical robotic platforms. However, physical hardware experiments are prohibitively expensive, time-consuming, and incapable of exhaustively evaluating the vast space of feasible states and boundary conditions. Simulation-based testing offers a scalable, repeatable framework for evaluating closed-loop system dynamics and detecting safety violations. Nevertheless, the validity of simulation-derived evidence critically depends on which inputs are selected and how thoroughly they cover the underlying input domain~\cite{bresolin2015surgery, araujo2023testing, voxel-lin_simulation-based_2025}.

Remote-Center-of-Motion (RCM) constraints serve as a fundamental mechanism to ensure a surgical instrument remains aligned with a fixed anatomical entry port. Systematic safety testing of an RCM-constrained surgical robot requires both a valid test domain and an effective testing strategy. The domain defines which physical configurations constitute distinct test inputs, whereas the testing strategy determines how a finite simulation budget is partitioned\textemdash whether toward broad coverage, unguided exploration, or focused evaluation of high-risk operational regions.

 Because motor commands directly govern robot actuation, the Cartesian product of valid joint parameters constitutes an input domain for test case generation. Nevertheless, this joint-space domain does not reflect task-level semantics.  
Independently selected motor commands can violate
coupled kinematic constraints, distinct  motor commands can represent the same
task-relevant state, and Euclidean distance between motor vectors does not reflect 
whether the corresponding tool poses exercise different safety conditions.
Another common test domain is the tool pose in Cartesian space, which introduces a complementary drawback: it describes spatial locations without guaranteeing that they are task-compatible or kinematically realizable by the robot. Consequently, exhaustive or uniform sampling across either domain risks expending substantial test budget on unreachable or redundant inputs while leaving critical, feasible task states unexplored.

This paper encodes test inputs using physically meaningful coordinates of
a \textit{task manifold}, which describes states satisfying structural task
constraints. Workspace and robot constraints further restrict these
coordinates to a \textit{feasible test domain}. Selected inputs are mapped
to motor commands for simulation, and their outcomes are associated with
the inputs' reference-grid cells for coverage and failure analysis.
Consequently, the test domain establishes the boundaries of valid execution and the reference framework for evaluation, while the testing strategy independently governs the allocation of the simulation budget across this domain.

We evaluate the method on an RCM-constrained surgical needle-withdrawal
task. Each test case specifies an initial needle configuration from which the
robot must withdraw through a fixed entry port, respecting the RCM
constraint~\cite{aghakhani2013rcm,taylor2003medical}. The initial task state is fully parameterized by two orientation variables and an insertion depth, subject to physical workspace, inverse-kinematic, and motor-limit constraints. These three task coordinates directly substitute for the raw motor variables of the robot under evaluation.

We compare three input representations---Cartesian XYZ, our
task-coordinate representation with feasibility filtering (labeled as Manifold), and
Principal Component Analysis (PCA)~\cite{abdi2010principal}---under random, traversal, and genetic search. This setup decouples valid input generation from search-driven failure discovery and coverage. We further classify failing trajectories by their first safety violation and the preceding geometric states to examine which failure behaviors each representation reveals.
Subsequently, a comparative evaluation between a defective and a patched controller assesses practical defect detection capabilities, followed by a MoveIt~2 case study to verify the transferability of our framework.

In summary, the paper makes three contributions:
\begin{itemize}[leftmargin=*]
  \item We introduce a task domain modeling method that embeds task-space and kinematic constraints directly into a minimal task-coordinate representation, eliminating invalid test candidates and maximizing execution throughput under a fixed sampling budget. 
  \item We propose a  testing framework that explicitly decouples the generation of valid test domains from the search strategies used to explore them, enabling independent analysis of input validity, spatial coverage, and failure detection. 
  \item We evaluate the Manifold representation against Cartesian XYZ and PCA baselines, showing its benefits for feasible test generation, systematic reference-cell coverage, and failure-pattern diversity. We further assess practical defect detection through controller variants and workflow transferability through a MoveIt~2~\cite{coleman2014moveit} planning pipeline.
\end{itemize}

\section{Related Work}
\label{sec:related-works}

\paragraph{Simulation-based testing of robotic and autonomous systems.}
A systematic review by Araujo et al.~\cite{araujo2023testing} identifies model checking, theorem proving, runtime verification, scenario-based testing, and simulation as primary paradigms for robotic software validation, while observing that test adequacy is frequently underreported or inconsistently defined. In system-level validation, Lin et al.~\cite{voxel-lin_simulation-based_2025} develop a multi-layered framework for autonomous surgical needle insertion, coupling a test generator and runtime monitor with simulation, physical hardware, and a digital twin.  Our work complements this system-level
testing perspective by constructing an RCM-derived feasible test domain  and evaluating the coverage efficiency of distinct search-space representations and testing strategies.

\paragraph{RCM-constrained surgical robotics and safety verification.}
Existing literature primarily leverages RCM constraints for controller design, inverse kinematics, and manipulability optimization~\cite{aghakhani2013rcm,taylor2003medical,colan2024manipulability}. While these approaches focus on \textit{executing} constrained surgical motions, our work addresses the complementary question: how systematically has the resulting controller been evaluated over its feasible input domain? On a closely related safety topic, Bresolin et al.~\cite{bresolin2015surgery} analyzed an autonomous puncturing task by modeling the controller and its associated uncertainty as a hybrid automaton, applying formal reachability analysis to verify the safety of planned subtask sequences. Whereas their method yields model-level guarantees for an abstract hybrid system, our approach directly exercises an executable robotic software stack alongside a physical safety oracle, using grid coverage to quantify the extent to which feasible RCM motions are tested.

\paragraph{Search-based testing and falsification.}
Search-based approaches allocate a limited simulation budget toward critical
or diverse scenarios. AmbieGen uses genetic and multi-objective search to
generate fault-revealing virtual environments for autonomous robots and
lane-keeping systems, optimizing both scenario difficulty and diversity
\cite{humeniuk2022ambiegen}. Tuncali et al. combine covering arrays with
simulation-based falsification to test autonomous vehicles containing machine
learning components \cite{tuncali2018simulation}. VerifAI provides a
general scenario-based framework for formal analysis, falsification, and
counterexample generation for AI-enabled systems \cite{dreossi2019verifai}.
These techniques demonstrate the value of optimization and formal monitors,
but typically assume that a suitable box of scenario parameters or a
domain-specific scenario description has already been defined. Our contribution
is complementary: it constructs the feasible testing domain before applying a
sampler or optimizer, and evaluates random, genetic, and systematic traversal
strategies against the same task-geometric reference.

\paragraph{Test domain representations and   coverages.}
Recent work shows that test effectiveness depends not only on the search
algorithm but also on the representation being searched. RILaST learns a
variational-autoencoder latent representation of autonomous-system scenarios
and performs evolutionary search in that latent space
\cite{humeniuk2025rilast}. DeepHyperion instead uses illumination search to
populate a behavior-feature map with diverse failure-inducing inputs
\cite{zohdinasab2021deephyperion}, and CIT4DNN applies combinatorial testing to
a generative model's latent space to target diverse and rare feature
combinations \cite{dola2024cit4dnn}. These learned or user-defined
representations are conceptually close to our decision to avoid the raw motor
space. However, their latent coordinates do not necessarily have direct
physical meaning or guarantee that decoded inputs satisfy task constraints.
Our task coordinates are derived explicitly from the RCM kinematics.
They have direct physical meanings, and feasibility checks retain only
parameter combinations that the robot can execute.

Coverage-oriented work further motivates treating testing as characterization
of a domain rather than only as counterexample discovery. GUARD models pass,
fail, and unknown regions of an autonomous-vehicle parameter space without
exhaustive grid discretization \cite{tu2023guard}. Wilbrandt and Dillmann
explore robot states satisfying motion constraints to estimate continuous surface coverage for
manufacturing tasks \cite{wilbrandt2026constraint}. The latter is geometrically
close to our approach, but its coverage denotes task reachability rather than
test adequacy and it does not include a failure oracle. Our work combines these
two perspectives: the task manifold describes constrained task states, feasibility checks identify
which states are executable, and test
coverage and failure-bearing cells summarize the evidence collected from the
executable system over that feasible domain.

\section{Background}

\subsection{RCM-Constrained Surgical Robot}
\label{sec:rcm-robot-background}

An RCM-constrained surgical robot positions an
instrument through a minimally invasive entry port. As illustrated in
\autoref{fig:rendered-robot}, the instrument has an elongated \emph{shaft}
and a \emph{tip}. The entry port defines a fixed pivot, the \emph{RCM point},
through which the shaft must pass while the instrument operates inside the
target organ (as shown in \autoref{fig:task-manifold-geometry}). The instrument can change orientation about this pivot or
translate along its shaft to insert or withdraw the tip.

With the fixed RCM point, the initial direction of the shaft  and  position of the tip can
be represented by three task parameters,
$\omega=(\theta_1,\theta_2,t)$~\cite{aghakhani2013rcm}.
As illustrated in 
\autoref{fig:rcm-parameter-effects}, the angles $\theta_1$ and $\theta_2$ describe rotations about the $x$- and
$y$-axes, respectively, and $t$ denotes the insertion depth along the shaft.
Changing either angle pivots the shaft about the RCM and moves the tip along
an arc at a fixed insertion depth. Changing $t$ moves the tip along the
shaft without changing its direction.

\begin{figure}[t]
  \centering
  \begin{subfigure}[t]{0.43\linewidth}
    \centering
    \includegraphics[width=0.85\linewidth]{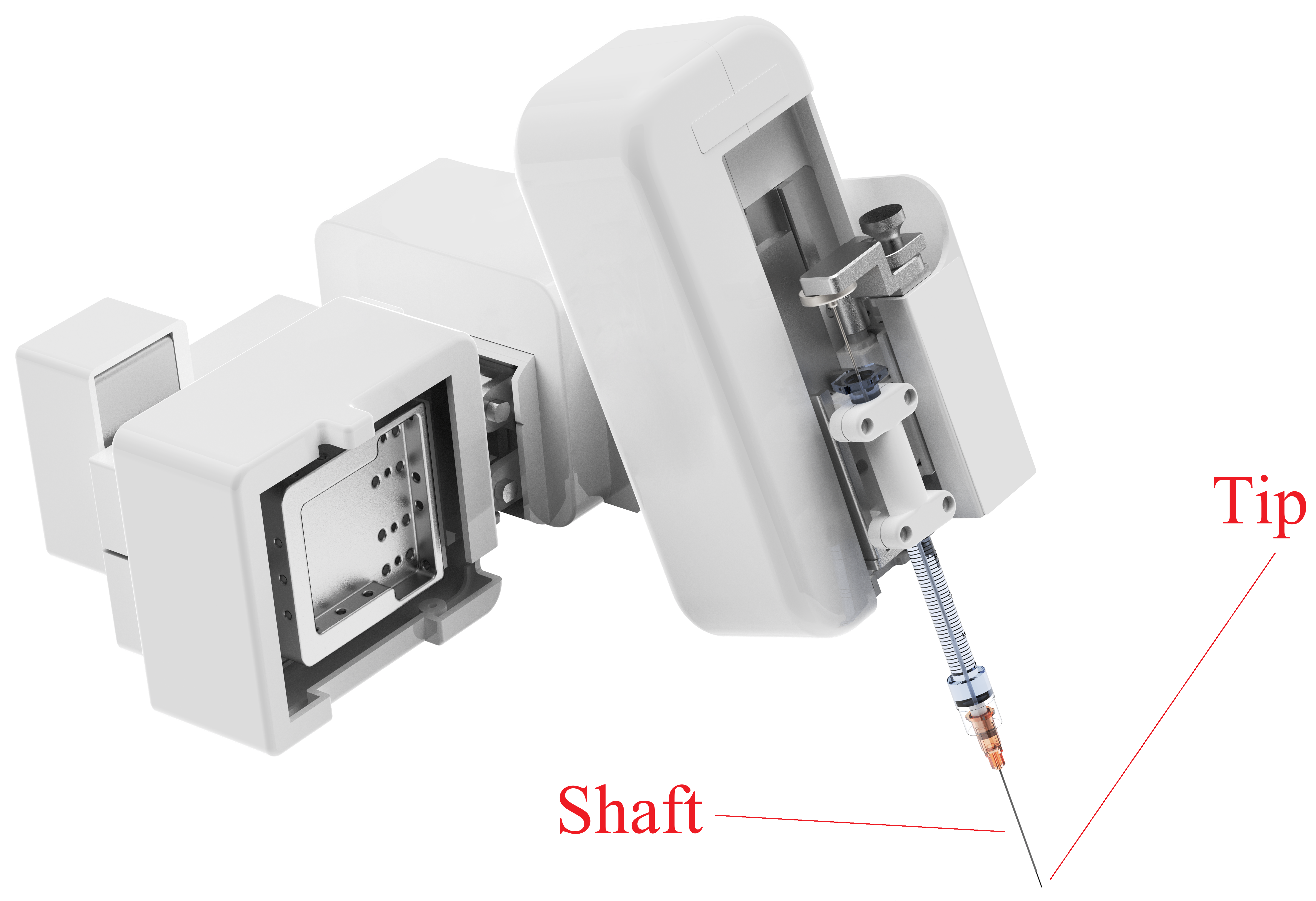}
    \caption{The ophthalmic surgical robot.}
    \label{fig:rendered-robot}
  \end{subfigure}\hfill
  \begin{subfigure}[t]{0.55\linewidth}
    \centering
    \includegraphics[width=\linewidth]{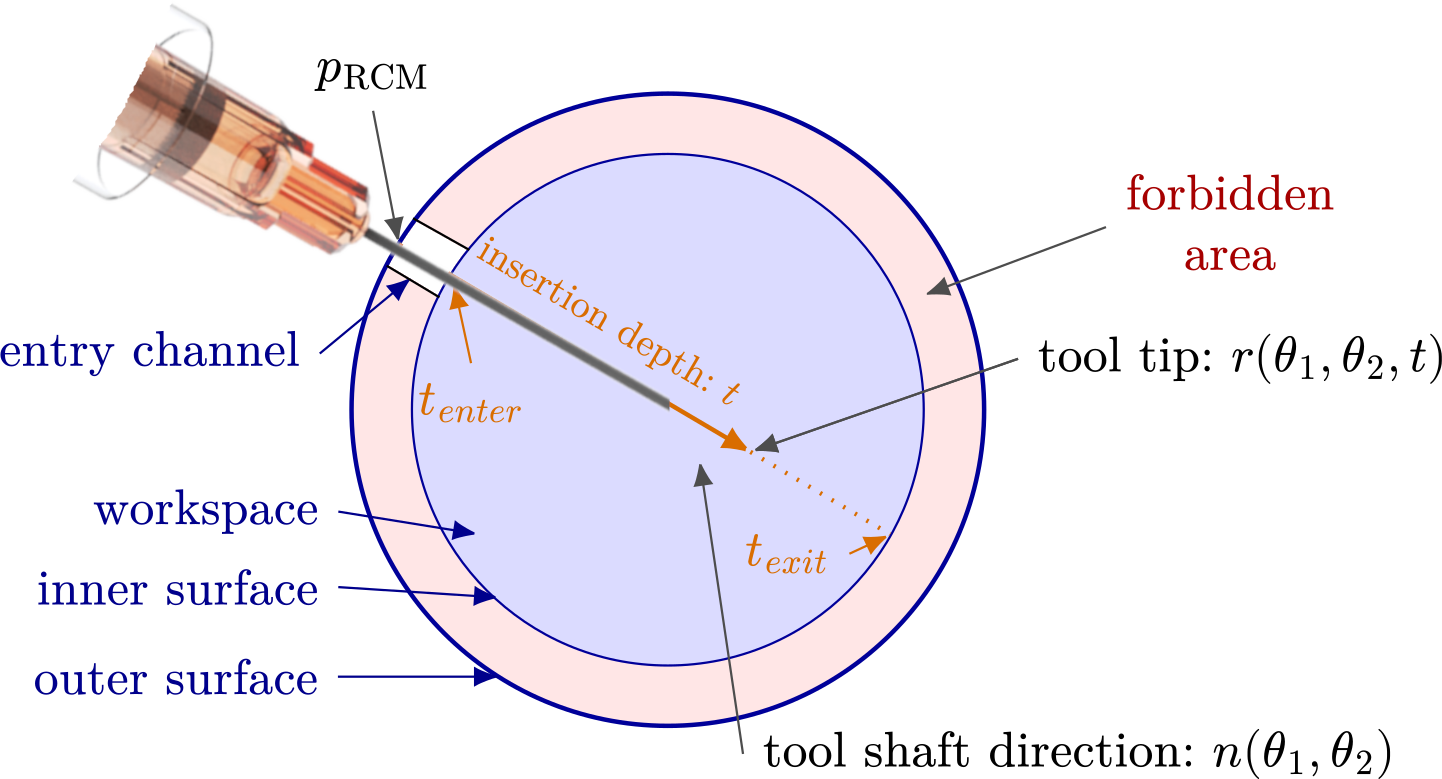}
    \caption{RCM-constrained task geometry.}
    \label{fig:task-manifold-geometry}
  \end{subfigure}
  \caption{The ophthalmic surgical robot and the initial-state geometry of the needle-withdrawal running example. The angles $(\theta_1,\theta_2)$ select the initial shaft direction, and $t$ selects the initial insertion depth along it. A test input is feasible only if it satisfies the task constraints and the robot can realize the specified configuration.}
  \Description{Left: a rendered view of the ophthalmic surgical robot. Right: the tool shaft passes through the fixed RCM point along a ray determined by the two angles. The insertion depth places the tip inside the blue admissible workspace; the surrounding red area is forbidden.}
\end{figure}
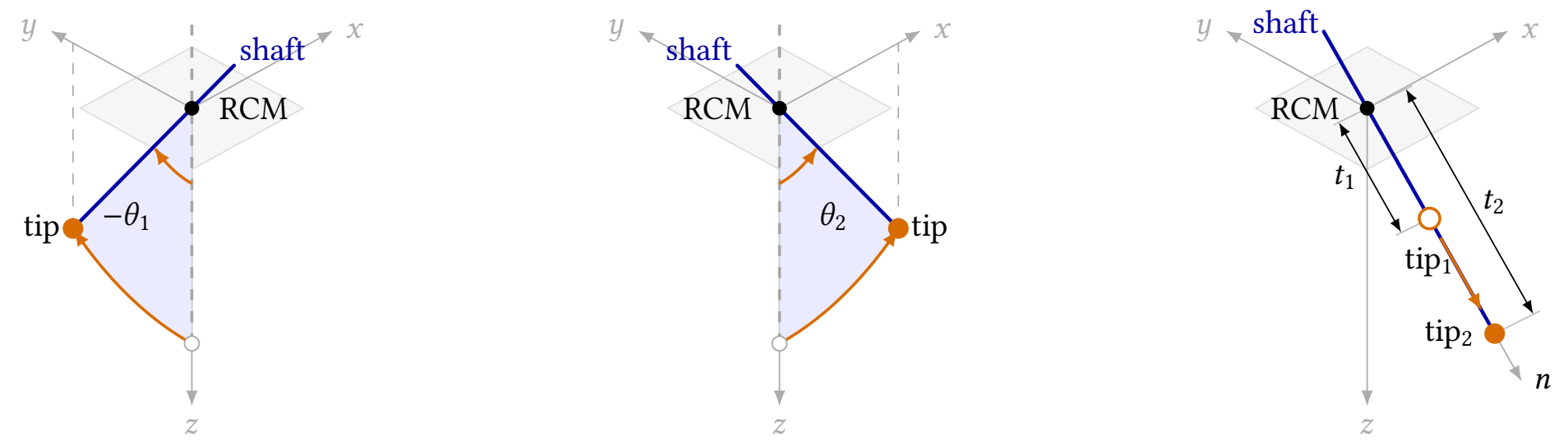
\begin{figure}[t]
  \centering
  \tikzset{
    rcm view/.style={x={(0.82cm,0.45cm)}, y={(-0.82cm,0.45cm)},
      z={(0cm,-1cm)}, font=\small},
    rcm axis/.style={gray!65, thin, -{Latex[length=1.5mm]}},
    rcm shaft/.style={blue!65!black, line width=1.0pt, line cap=round},
    rcm reference/.style={gray!70, thick, dashed},
    rcm motion/.style={orange!85!black, thick, -{Latex[length=1.7mm]}},
    rcm dimension/.style={thin, {Latex[length=1.3mm]}-{Latex[length=1.3mm]}}
  }
  \newcommand{\rcmcoordinateframe}{%
    \path[use as bounding box] (-1.65cm,-3.05cm) rectangle (2.05cm,1.25cm);
    \coordinate (O) at (0,0,0);
    \filldraw[fill=gray!7, draw=gray!25, thin]
      (-0.65,-0.65,0) -- (0.65,-0.65,0) --
      (0.65,0.65,0) -- (-0.65,0.65,0) -- cycle;
    \draw[rcm axis] (O) -- (1.65,0,0) node[right] {$x$};
    \draw[rcm axis] (O) -- (0,1.65,0) node[left] {$y$};
    \draw[rcm axis] (O) -- (0,0,2.85) node[below] {$z$};
  }
  \begin{minipage}[t]{0.32\linewidth}
    \centering
    \resizebox{0.7\linewidth}{!}{%
    \begin{tikzpicture}[rcm view]
      \rcmcoordinateframe
      \coordinate (P) at (0,{2.25*sin(38)},{2.25*cos(38)});
      \path[fill=blue!8] (O) -- (0,0,2.25)
        -- plot[domain=0:38, samples=25, variable=\a]
          (0,{2.25*sin(\a)},{2.25*cos(\a)}) -- cycle;
      \draw[gray!65, thin, dashed] (P) -- (0,{2.25*sin(38)},0);
      \draw[rcm reference] (0,0,-0.8) -- (0,0,2.25);
      \draw[rcm motion] plot[domain=0:38, samples=25, variable=\a]
        (0,{2.25*sin(\a)},{2.25*cos(\a)});
      \draw[rcm shaft] (0,{-0.8*sin(38)},{-0.8*cos(38)}) -- (P);
      \node[blue!65!black, above right] at
        (0,{-0.65*sin(38)},{-0.65*cos(38)}) {shaft};
      \draw[rcm motion] plot[domain=0:38, samples=20, variable=\a]
        (0,{0.72*sin(\a)},{0.72*cos(\a)});
      \node[below left] at (0,{0.96*sin(19)},{0.96*cos(19)}) {$-\theta_1$};
      \draw[gray!70, fill=white] (0,0,2.25) circle[radius=2pt];
      \fill[orange!85!black] (P) circle[radius=2.8pt]
        node[left, text=black] {tip};
      \fill (O) circle[radius=2pt] node[right=4pt] {RCM};
    \end{tikzpicture}%
    }
    \par\smallskip
    {\small (a) $\theta_1$: rotate about x-axis}
  \end{minipage}\hfill
  \begin{minipage}[t]{0.32\linewidth}
    \centering
    \resizebox{0.7\linewidth}{!}{%
    \begin{tikzpicture}[rcm view]
      \rcmcoordinateframe
      \coordinate (P) at ({2.25*sin(38)},0,{2.25*cos(38)});
      \path[fill=blue!8] (O) -- (0,0,2.25)
        -- plot[domain=0:38, samples=25, variable=\a]
          ({2.25*sin(\a)},0,{2.25*cos(\a)}) -- cycle;
      \draw[gray!65, thin, dashed] (P) -- ({2.25*sin(38)},0,0);
      \draw[rcm reference] (0,0,-0.8) -- (0,0,2.25);
      \draw[rcm motion] plot[domain=0:38, samples=25, variable=\a]
        ({2.25*sin(\a)},0,{2.25*cos(\a)});
      \draw[rcm shaft] ({-0.8*sin(38)},0,{-0.8*cos(38)}) -- (P);
      \node[blue!65!black, above left] at
        ({-0.65*sin(38)},0,{-0.65*cos(38)}) {shaft};
      \draw[rcm motion] plot[domain=0:38, samples=20, variable=\a]
        ({0.72*sin(\a)},0,{0.72*cos(\a)});
      \node[below right] at ({0.96*sin(19)},0,{0.96*cos(19)}) {$\theta_2$};
      \draw[gray!70, fill=white] (0,0,2.25) circle[radius=2pt];
      \fill[orange!85!black] (P) circle[radius=2.8pt]
        node[right, text=black] {tip};
      \fill (O) circle[radius=2pt] node[left=4pt] {RCM};
    \end{tikzpicture}%
    }
    \par\smallskip
    {\small (b) $\theta_2$: rotate about y-axis}
  \end{minipage}\hfill
  \begin{minipage}[t]{0.32\linewidth}
    \centering
    \resizebox{0.7\linewidth}{!}{%
    \begin{tikzpicture}[rcm view]
      \rcmcoordinateframe
      \pgfmathsetmacro{\nx}{cos(20)*sin(18)}
      \pgfmathsetmacro{\ny}{-sin(20)}
      \pgfmathsetmacro{\nz}{cos(20)*cos(18)}
      \coordinate (Pone) at ({1.15*\nx},{1.15*\ny},{1.15*\nz});
      \coordinate (Ptwo) at ({2.35*\nx},{2.35*\ny},{2.35*\nz});
      \draw[rcm axis] (O) -- ({2.85*\nx},{2.85*\ny},{2.85*\nz})
        node[right, text=black] {$n$};
      \draw[rcm shaft] ({-0.8*\nx},{-0.8*\ny},{-0.8*\nz}) -- (Ptwo);
      \node[blue!65!black, above left] at
        ({-0.65*\nx},{-0.65*\ny},{-0.65*\nz}) {shaft};
      \draw[gray!50, thin] (O) -- ($(O)+(-0.32cm,-0.16cm)$);
      \draw[gray!50, thin] (Pone) -- ($(Pone)+(-0.32cm,-0.16cm)$);
      \draw[rcm dimension] ($(O)+(-0.27cm,-0.135cm)$) --
        ($(Pone)+(-0.27cm,-0.135cm)$) node[midway, left] {$t_1$};
      \draw[gray!50, thin] (O) -- ($(O)+(0.43cm,0.215cm)$);
      \draw[gray!50, thin] (Ptwo) -- ($(Ptwo)+(0.43cm,0.215cm)$);
      \draw[rcm dimension] ($(O)+(0.37cm,0.185cm)$) --
        ($(Ptwo)+(0.37cm,0.185cm)$) node[midway, right] {$t_2$};
      \draw[rcm motion] ({1.35*\nx},{1.35*\ny},{1.35*\nz}) --
        ({2.12*\nx},{2.12*\ny},{2.12*\nz});
      \draw[orange!85!black, thick, fill=white] (Pone) circle[radius=2.8pt]
        node[below=4pt, text=black] {tip$_1$};
      \fill[orange!85!black] (Ptwo) circle[radius=2.8pt]
        node[left=2pt, text=black] {tip$_2$};
      \fill (O) circle[radius=2pt] node[left=4pt] {RCM};
    \end{tikzpicture}%
    }
    \par\smallskip
    {\small (c) $t$: insertion depth}
  \end{minipage}
  \caption{Illustrations of the RCM task parameters.}
  \Description{Three diagrams share a three-dimensional coordinate frame
  with positive z pointing downward. Each shaft extends from above the
  RCM to a tip below it. The first panel shows negative theta one tilting
  the shaft toward positive y, to the left in this view. The second shows
  positive theta two tilting it toward positive x, to the right. Their tip
  arcs lie in the yz and xz planes, respectively. The third panel
  shows two tip positions at different depths on the same downward shaft.}
  \label{fig:rcm-parameter-effects}
\end{figure}

\subsection{Task Manifold}

Although $(\theta_1,\theta_2,t)$ directly specifies the initial shaft orientation and tip position, the RCM constraint intrinsically couples the robot's joint variables. Consequently, these three independent task coordinates form a local parameterization of all RCM-aligned configurations. Viewing each valid coordinate triplet as a task state exemplifies the concept of a \emph{manifold}—a topological space that locally admits a coordinate system of fixed dimension. We formally define the space of task states satisfying these structural constraints as the \emph{task manifold}.

Compared with motor-command and Cartesian parameter spaces, which do not inherently enforce RCM alignment, the task manifold incorporates this constraint into its structure, so that every represented state directly satisfies the required alignment between the shaft and the fixed entry point.

\subsection{Motivating Example: Needle Withdrawal}\label{sec:example}
We use an ophthalmic needle-withdrawal task as a motivating example throughout
the paper:  
A needle tip positioned inside the eye must perform 
a linear withdrawal along its initial shaft direction to a target outside
the eye. As illustrated in \autoref{fig:task-manifold-geometry}, the planned trajectory guides the instrument tip through the admissible entry channel while avoiding the surrounding tissue, which is modeled as a forbidden region. 
During operation, the controller of the robot
 generates and tracks the trajectory originating from a given task state. However, due to control tracking errors or scheduling latencies, the resulting motion may deviate into the forbidden region or penetrate the eye boundary outside the designated entry port. 

\begin{figure}[t]
\centering
\begin{minipage}[t]{0.49\linewidth}
  \centering
  \includegraphics[width=0.7\linewidth]{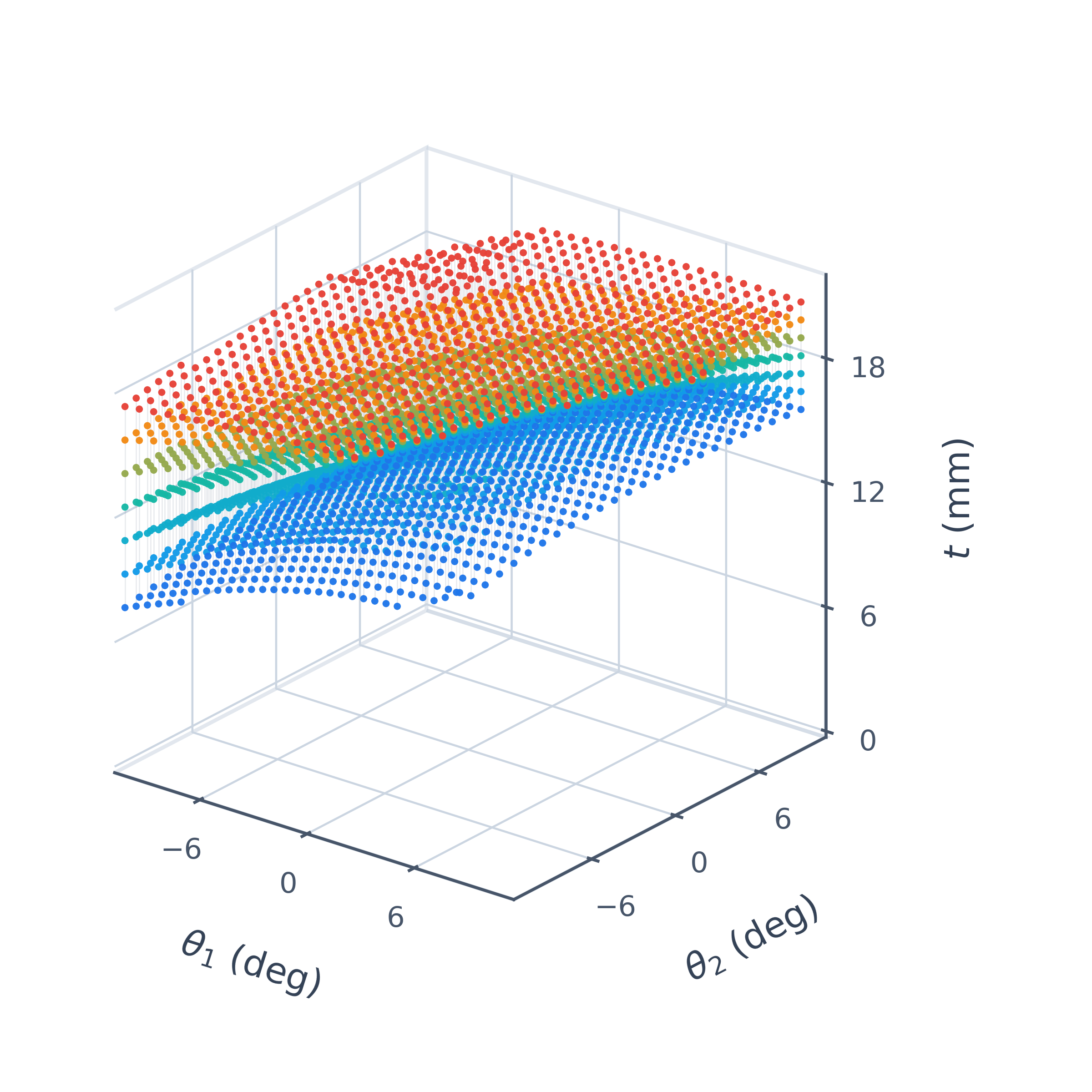}

  \small (a) Points in task coordinates $(\theta_1,\theta_2,t)$
\end{minipage}\hfill
\begin{minipage}[t]{0.49\linewidth}
  \centering
  \includegraphics[width=0.7\linewidth]{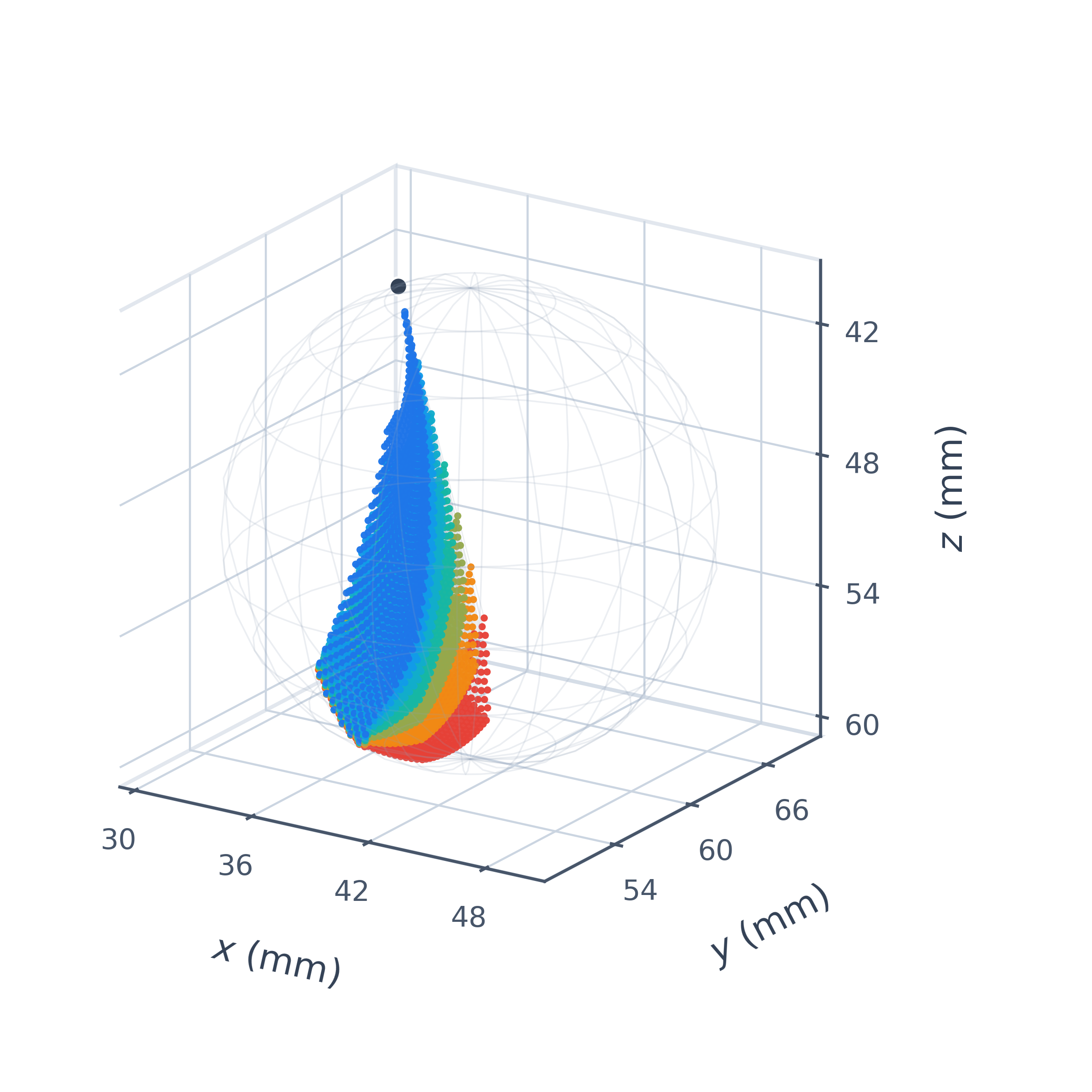}

  \small (b) Points in Cartesian coordinates $(x,y,z)$
\end{minipage}
\caption{The same feasible initial states of the withdrawal running example
represented in task coordinates and Cartesian tool-tip coordinates.}
\label{fig:manifold-coordinate-comparison}
\Description[The same feasible points in task and Cartesian coordinates]{%
The left panel plots the feasible points by two shaft angles and insertion
depth, where they form vertical depth sequences.  The right panel plots the
same points by Cartesian tool-tip position inside a spherical workspace, where
the sequences spread outward from the RCM point and form a curved, nonuniform
cloud.  Corresponding depth layers use matching colors in both panels.
}
\end{figure}

\autoref{fig:manifold-coordinate-comparison} shows identical initial candidate states of the running example in both task coordinates and Cartesian tool-tip
coordinates, with matching colors for corresponding depth layers.
States sharing a common shaft orientation form linear vertical sequences along $t$ in task space, whereas they project as divergent rays originating from the RCM point in Cartesian space. %

\section{Approach}
\label{sec:approach}

\subsection{Overview}

\begin{figure}[!t]
\centering
\includegraphics[width=0.8\textwidth]{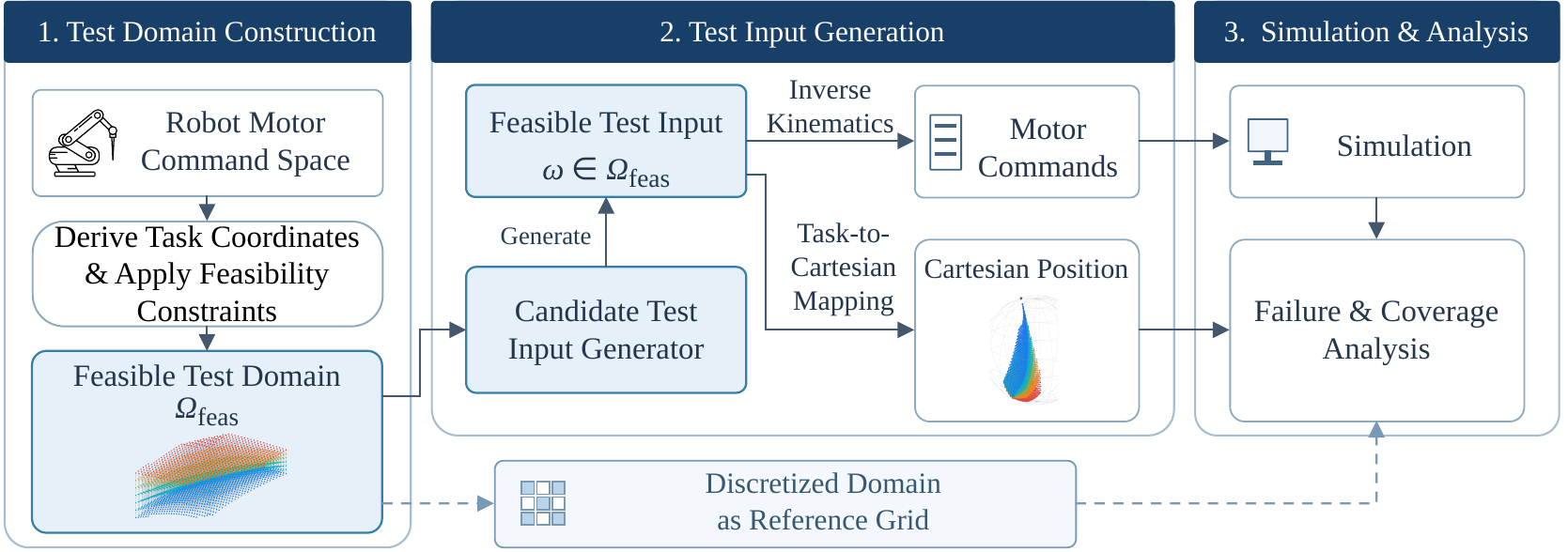}
\caption{Overview of testing with a feasible test domain.}
\label{fig:method-overview}
\Description[Overview of testing with feasible test domain]{%
Robot analysis identifies a feasible test domain from the
motor-command space and task constraints.  A candidate generator uses this domain to produce a valid test
input.  The input is converted to motor commands for simulation and to a
Cartesian position; the simulation result and Cartesian position are then used
for failure and coverage analysis.
}
\end{figure}

Our framework provides systematic end-to-end testing for robotic systems executing defined motion tasks, encompassing their kinematic models, motion planners, and low-level controllers.

As illustrated in \autoref{fig:method-overview}, the framework operates in three sequential stages. First, in test domain construction, we derive the feasible test domain from bounded motor commands by applying robot kinematics and physical operational constraints. Second, in test input generation, a candidate generator samples valid task-coordinate inputs, and transforms them to executable motor commands and Cartesian tool-tip initial states for simulation. Third, in simulation and analysis, the simulator executes the resulting trajectory; a safety oracle classifies the trial outcome; and the input is mapped onto the reference grid to evaluate spatial coverage and delineate failure regions.

\subsection{Construct a Feasible Test Domain from Motor Commands}
\label{sec:approach-construct-feasible-domain}

We first define the motor-command and task-coordinate domains alongside the conditions for test-input feasibility, and then apply this formulation to our motivating example.

Let $q=(q_1,\ldots,q_m)^\top \in \mathbb{R}^m$ be a command vector for $m$ motors, where $\mathbb{R}$ represents the set of real numbers.
Given the component-wise bounds $[\underline{q}_i, \overline{q}_i]$ for each motor $i \in \{1, \ldots, m\}$, the motor-command space $\mathcal{Q}$ is defined by the Cartesian product:
\begin{equation}
  \mathcal{Q}
  = \prod_{i=1}^{m}[\underline q_i,\overline q_i]
  \subseteq \mathbb{R}^{m}.
\end{equation}
These bounds constrain individual motor commands but do not ensure that
the resulting tool shaft direction and tool tip location satisfy task constraints, such as
RCM alignment and workspace boundaries.

To express these constraints directly and reduce the input dimensionality,
we parameterize test inputs using $d$ independent task coordinates
$\omega\in\Omega\subseteq\mathbb{R}^d$, where $\Omega$ is the candidate
task-coordinate domain before feasibility checks.
For the motivating example, the test %
input $\omega=(\theta_1,\theta_2,t)$ specifies
the initial shaft direction and insertion depth introduced in
\autoref{sec:rcm-robot-background}. This parameterization satisfies the RCM
alignment constraint by construction.

Two predicates determine whether a test input $\omega$ is admissible:
$C_{\mathrm{task}}(\omega)$ verifies workspace and task-level safety
requirements, while $C_{\mathrm{robot}}(q,\omega)$ checks whether a motor
command $q\in\mathcal{Q}$ realizes $\omega$ under robot-specific kinematic
and safety constraints. Task states satisfying these requirements and
admitting at least one valid motor realization are the \emph{feasible task
states}. The task-coordinate vectors representing them form the
\emph{feasible test domain}:
\begin{equation}
  \Omega_{\mathrm{feas}}
  = \left\{
    \omega\in\Omega \;\middle|\;
    C_{\mathrm{task}}(\omega)
    \land \exists q\in\mathcal{Q}: C_{\mathrm{robot}}(q,\omega)
  \right\}.
\end{equation}
These checks ensure that an initial %
configuration is admissible
and realizable.
Candidate inputs are selected from $\Omega_{\mathrm{feas}}$ and mapped to
valid motor commands for execution.
\autoref{tab:task-space-terminology} summarizes the domains and their roles.

\begin{table}[htbp]
\centering
\caption{Terminology for motor commands and task-based test inputs.}
\label{tab:task-space-terminology}
\small
\begin{tabular}{@{}llp{0.6\linewidth}@{}}
\toprule
Term & Notation & Meaning \\
\midrule
Task manifold & --- &
Task states satisfying structural task constraints, such as RCM alignment. \\
Motor-command space & $\mathcal{Q}$ &
Motor-command vectors satisfying individual motor limits. These bounds alone
do not ensure task feasibility. \\
Task-coordinate domain & $\Omega$ &
Candidate task-coordinate vectors before feasibility checks.
For the motivating example, it is $(\theta_1,\theta_2,t)$. %
\\
Feasible test domain & $\Omega_{\mathrm{feas}}$ &
The subset of $\Omega$ satisfying task constraints and admitting at least one
valid motor realization in $\mathcal{Q}$. Sampling and search operate on this domain. \\
\bottomrule
\end{tabular}
\end{table}

We now instantiate this construction for the motivating example:
we first map its task coordinates to the initial needle geometry, then
determine which coordinate combinations are feasible.

\subsubsection{Task Coordinates}

For an initial configuration $\omega=(\theta_1,\theta_2,t)$, the two angles
determine the shaft direction:
\begin{equation}
  n(\theta_1,\theta_2)
  =
  \begin{bmatrix}
    \cos\theta_1 \sin\theta_2\\
    -\sin\theta_1\\
    \cos\theta_1 \cos\theta_2
  \end{bmatrix}.
\end{equation}

The corresponding 3D Cartesian tip position is
\begin{equation}
  r(\theta_1,\theta_2,t)
  =
  p_{\mathrm{RCM}} + t\,n(\theta_1,\theta_2),
  \qquad \textrm{where }t \geq 0.
\end{equation}

Fixing $(\theta_1,\theta_2)$ selects an RCM-aligned ray, and varying $t$
places the initial needle tip along it
(\autoref{fig:task-manifold-geometry}). We next determine which depths
along this ray satisfy the task and robot constraints.

\subsubsection{Constraints and Feasible Region}

For a fixed shaft direction, feasible initial depths must satisfy
both the task geometry and robot realizability constraints. We first
identify the depths allowed by the task geometry:
\begin{equation}
  T_{\mathrm{task}}(\theta_1,\theta_2)
  =
  \left\{t\geq0\;\middle|\;
  C_{\mathrm{task}}\bigl((\theta_1,\theta_2,t)^\top\bigr)\right\}.
\end{equation}
Here $C_{\mathrm{task}}$ requires the shaft ray determined by
$n(\theta_1,\theta_2)$ to pass through the permitted entry channel centered at
$p_{\mathrm{RCM}}$ and the tip position $r(\theta_1,\theta_2,t)$ to lie inside
the admissible workspace. If the ray enters through the permitted channel, the
admissible depths form a segment inside the target organ, from the RCM entry point to
the opposite inner surface. Hence, when such a segment exists, we have 
\begin{equation}
  T_{\mathrm{task}}(\theta_1,\theta_2)
  =
  [t_{\mathrm{enter}}(\theta_1,\theta_2),
   t_{\mathrm{exit}}(\theta_1,\theta_2)].
\end{equation}
Here $t_{\mathrm{enter}}=0$ is the depth at the RCM entry point, and
$t_{\mathrm{exit}}$ is the depth at the opposite inner surface. We obtain the
endpoints ($t_{\mathrm{enter}}$ and $t_{\mathrm{exit}}$) by intersecting the ray %
with the spherical workspace boundary. If the ray has no admissible inward segment,
$T_{\mathrm{task}}(\theta_1,\theta_2)=\varnothing$.

Task-admissible depths must also be realizable by the robot. We define the
robot-admissible depth set by requiring at least one motor command
$q\in\mathcal{Q}$ that realizes the corresponding initial configuration
while satisfying the robot's kinematic and safety constraints:
\begin{equation}
  T_{\mathrm{robot}}(\theta_1,\theta_2)
  =
  \left\{t\geq0\;\middle|\;
  \exists q\in\mathcal{Q}:
  C_{\mathrm{robot}}(q,(\theta_1,\theta_2,t)^\top)\right\}.
\end{equation}
For the robot under evaluation, these checks filter out kinematic singularities and numerically unstable inverse solutions that could induce erratic control outputs, compromised system robustness, or unsafe trajectory execution. Furthermore, these evaluations enforce physical actuator limits and reject task inputs whose reconstructed tool shaft violates RCM alignment tolerances. Intersecting the task-level and robot-level valid sets yields the feasible depth interval for a given orientation: 
\begin{equation}
  T(\theta_1,\theta_2)
  =
  T_{\mathrm{task}}(\theta_1,\theta_2)
  \cap
  T_{\mathrm{robot}}(\theta_1,\theta_2).
\end{equation}

For $\omega=(\theta_1,\theta_2,t)^\top\in\Omega$, the task predicate in the
general definition is equivalent to $t\in T_{\mathrm{task}}(\theta_1,\theta_2)$,
and the existence of a valid motor command is equivalent to
$t\in T_{\mathrm{robot}}(\theta_1,\theta_2)$. The motor limits and the
robot's coordinate conversion bound the candidate orientations in $\Omega$.
For each candidate orientation, the task and robot checks determine
$T(\theta_1,\theta_2)$; an orientation is feasible exactly when this set is
nonempty. Collecting the feasible depths for these orientations gives the same
domain:
\begin{equation}
\Omega_{\mathrm{feas}}
=
\left\{
  (\theta_1,\theta_2,t)^\top\in\Omega
  \;\middle|\;
  t \in T(\theta_1,\theta_2)
\right\}.
\end{equation}
Each point in $\Omega_{\mathrm{feas}}$ possesses a well-defined physical interpretation—a unique tool shaft orientation and a corresponding tip position along its RCM directional ray—guaranteeing a valid motor realization prior to simulation. Physical robot kinematic properties determine the concrete numerical orientation bounds and define the robot-specific mapping.

\subsection{Input Generation}
\label{sec:input-generation}

Input generation selects test candidates directly from the feasible test domain derived during domain construction.  While a generation strategy governs which feasible states to evaluate and their execution order, the feasible domain dictates candidate admissibility. Specifically, the generator selects a task coordinate input $\omega\in\Omega_{\mathrm{feas}}$, which is mapped to an executable motor command $q=\Psi(\omega)$. The transformation 
$\Psi:\Omega_{\mathrm{feas}}\rightarrow\mathcal{Q}$ yields a joint command
that satisfies $C_{\mathrm{robot}}(q,\omega)$, which is then executed through the unified testing pipeline.

\subsection{Result Analysis}
\label{sec:approach-analysis}
For each executed test, we analyze both its initial configuration and the
resulting trajectory. In the withdrawal running example, the initial
needle configuration is mapped to a reference grid to measure which parts
of the feasible test domain have been explored. The recorded withdrawal
trajectory is checked by an offline safety oracle to determine whether
the run passes or fails. We first define the reference grid and coverage
metric, then specify the oracle.

\subsubsection{Reference Grid and Coverage}

In the running example, each reference cell groups initial needle
configurations with similar shaft directions and insertion depths.
To discretize the continuous feasible test domain, we uniformly partition
each orientation angle into grid steps. 
We retain angular coordinate pairs $(\theta_{1,i}, \theta_{2,j})$ that yield a non-empty feasible depth interval, or for which stochastic search identifies a valid orientation within the cell under a fixed attempt budget. For each retained orientation, its  feasible depth range is partitioned into $K$ equal-width bins.
Indexing the angular
coordinates by $i$ and $j$ and the depth bins by $k$, each tuple $(i,j,k)$
defines a reference cell $c_{ijk}\subset\Omega_{\mathrm{feas}}$.

Each executed input is expressed in task coordinates and mapped to
 at most one reference cell. 
 First, its orientation angles are quantized independently to the nearest discretized grid values. Inputs exceeding the angular bounds or landing on an infeasible orientation pair are discarded as unmapped. Otherwise, the recorded insertion depth is projected onto the corresponding directional depth interval and assigned to its nearest depth bin, yielding the cell assignment.

Let $\mathcal{S}$ denote the set of executed test inputs in
the evaluation, $\mathcal{C}$ represent the set of all valid reference cells and
$\mathcal{C}_{\mathrm{visited}}(\mathcal{S})$ be the set of cells
containing at least one mapped input from  $\mathcal{S}$.
A reference cell is counted as visited when an executed test starts from
a configuration mapped to it, regardless of whether the resulting
withdrawal passes or fails.
We define
\begin{equation}
  \operatorname{Coverage}(\mathcal{S})
  =
  \frac{|\mathcal{C}_{\mathrm{visited}}(\mathcal{S})|}
       {|\mathcal{C}|}.
\end{equation}

Coverage measures the proportion of valid reference cells assigned at
least one executed test.
While higher cell coverage indicates that the generated test suite explores a broader, more diverse region of the feasible task space, coverage alone is insufficient to guarantee controller safety. Consequently, we complement coverage with an offline trajectory safety oracle to evaluate dynamic execution behavior and detect safety violations.

\subsubsection{Trajectory Safety Oracle}

The safety oracle classifies simulation runs using execution status,
forbidden-region contacts, and boundary-crossing checks. A test fails if the tool tip breaches forbidden anatomical regions of the target organ or violates RCM constraints during trajectory execution.
For the withdrawal task, let the recorded tip trajectory be $\tau=(p_0,p_1,\ldots,p_T)$,
where $p_i\in\mathbb{R}^3$ is a sampled tip position.
Let $o$ denote the workspace center, $r_{\mathrm{in}}$ and $r_{\mathrm{out}}$ its
inner and outer radii, and $EC$ the permitted RCM entry channel. The forbidden
area is
\begin{equation}
  FS=\{p\mid r_{\mathrm{in}}\leq\|p-o\|\leq r_{\mathrm{out}}\} \setminus EC.
\end{equation}
The entry channel and forbidden area of the motivating example are illustrated in \autoref{fig:task-manifold-geometry}.

A point violation occurs when $\exists p_i\in FS$,
meaning the tip position lies inside the forbidden shell but outside the permitted entry channel. 

During withdrawal, because trajectories are sampled discretely, the distal tool tip could potentially penetrate and exit a forbidden region between two consecutive waypoints.
To detect unsampled boundary crossings,  the line-segment crossing predicate  $V_{\mathrm{cross}}(p_{i-1},p_i)$ evaluates to True if the linear segment connecting $p_{i-1}$ and $p_i$ transitions between the inner boundary $\|p-o\|<r_{\mathrm{in}}$  and
the outer boundary $\|p-o\| > r_{\mathrm{out}}$ without traversing the designated entry channel $EC$.
Formally, a simulation trajectory $\tau$ passes if and only if:
\begin{align}
Oracle(\tau)=\mathrm{pass}\iff
&\ \mathrm{ret}=0 \\
&\wedge\ \forall i,\ p_i\notin FS \\
&\wedge\ \forall i>0,\ \neg V_{\mathrm{cross}}(p_{i-1},p_i),
\end{align}
where $\mathrm{ret}=0$ indicates normal simulation completion without system exceptions. Consequently, simulation runtime crashes, forbidden-shell point contacts outside the entry channel, and discrete boundary segment violations are all classified as failures. 

\section{Research Questions}

We evaluate candidate validity, testing performance across spaces and
strategies, and practical defect detection and transferability through three
research questions.

\begin{description}[leftmargin=!,labelwidth=1.2cm]
  \item[RQ1] How effectively does the feasible test domain generate
  valid test cases, and how can this benefit testing?

  \item[RQ2] How does the constrained feasible test domain affect spatial coverage, failure discovery, and the diversity of observed failure behaviors across test generation strategies and domain representations?

  \item[RQ3] Can the method support defect detection and transfer across robot kinematics?  %
\end{description}

To answer $RQ1$, we compare random generation in Cartesian XYZ, Manifold, and PCA~\cite{abdi2010principal} spaces,
  counting valid executable cases and all candidate cases required to obtain the same number of 
  valid cases per space. We then count the
  proportion of valid inputs among the candidate inputs as the candidate validity.

To answer $RQ2$, we compare spatial coverage and failure discovery for a
five-motor controller across the three input spaces under random, traversal,
and genetic algorithm (GA) strategies, using the same simulator, safety
oracle, and reference-grid mapping. Random and GA each execute four runs of 4,603 tests per space. Traversal uses the same resolution to sample test inputs per space.
The budget of 4,603 is motivated by the zero-failure criterion: if all tests pass under independent, identically distributed sampling, we can conclude with $99\%$ confidence that the failure probability under that sampling distribution is below $0.1\%$.
We report reference-grid coverage and failure-pattern counts, and examine
spatial distributions. We distinguish the geometric
state sequences leading to the first safety violation as patterns, and use patterns to
classify the failure behaviors.

To answer $RQ3$, we compare two versions of controller implementations for a five-motor robot
with random and GA
  testing strategies, performing four runs of 4,603 tests for each controller--strategy pair.
  We examine both aggregate failures and failures near the motor-limit region
  that activates the defect. We then apply the workflow to a Kinova Gen3 arm
  in MoveIt~2~\cite{coleman2014moveit} implementing the same task, replacing analytic inverse kinematics with a numerical
  realization policy. %
  We use both systematic traversal and random testing to quantify motion-planning failures, dynamic trajectory violations, and cell coverage within the reference grid.

\section{Experimental Setup}

\subsection{Evaluation Subject}
\label{sec:evaluation-subject}

We use both the five-motor surgical robot and the Kinova Gen3 arm to perform the
withdrawal task introduced in \autoref{sec:example}, with the
withdrawal target placed $1$~mm beyond the outer sclerotomy boundary.
\autoref{tab:subject-parameters} lists the five-motor robot's task and
kinematic parameters, including the angular bounds obtained by the domain
construction in \autoref{sec:approach-construct-feasible-domain}.
Cartesian quantities use millimeters; motor values use native actuator units.

\begin{table}[!ht]
\centering
\caption{Robot and task parameters for the five-motor evaluation subject.}
\label{tab:subject-parameters}
\begin{tabular}{ll}
\toprule
Parameter & Value \\
\midrule
$q_1,\ldots,q_4$ limits & $[0,0.25]$ \\
$q_5$ limits & $[0.12154,0.30]$ \\
Workspace center & $(40.13382,59.36,49.98868)$ mm \\
RCM point & $(36.44,59.36,39.84)$ mm \\
Inner / outer workspace radius & $10.8$ mm / $12.0$ mm \\
Permitted entry-channel radius & $0.5$ mm \\
$\theta_1,\theta_2$ bounds & $[-47.386^\circ,47.386^\circ]$ \\
\bottomrule
\end{tabular}
\end{table}

\subsection{Mappings between Test Input and Motor Commands}

For the five-motor robot, analytic inverse kinematics implements $\Psi$ as
\begin{equation}
\Psi(\theta_1,\theta_2,t)
=
\bigl(a_1(\theta_1,\theta_2),\ldots,a_4(\theta_1,\theta_2),\,t+b(\theta_1,\theta_2)\bigr),
\end{equation}
where functions $a_1,\ldots,a_4$ and $b$ are uniquely determined by the robot's kinematic geometry.
The first four motor commands govern the spatial orientation, while the fifth command %
controls insertion depth.

The transformation is accepted only if it is nonsingular and numerically stable,
and the resulting command satisfies the motor limits and RCM spatial tolerances.

For the Kinova Gen3 arm in MoveIt~2, $\Psi$ converts task coordinates to a
tool pose and uses numerical IK with a fixed seed policy to select a
realization of the redundant arm. Configurations must pass checks of
forward-kinematic pose accuracy, motor limits, collisions, and initial
RCM alignment.

\subsection{Test Space Representations and Test Case Generation Strategies}

The comparative evaluation assesses three input space representations. Cartesian XYZ space parameterizes the 3D tool-tip position as $(x,y,z)$. The
task-coordinate representation with feasibility filtering (labeled Manifold in the results) parameterizes inputs via orientation angles and normalized
insertion depth $s\in[0,1]$, which linearly parameterizes $t$ over
the feasible insertion-depth interval for the selected orientation. The latent 
PCA space parameterizes candidate configurations in a reduced coordinate frame learned from valid motor poses. %
To isolate the motor input domain, this dimensionality reduction is performed following input normalization, without incorporating simulation outcomes or manifold samples. PCA candidates are subsequently 
mapped back to motor commands via the inverse transform. All candidate spaces enforce identical task-geometry, inverse-kinematic, and motor-limit criteria to decide whether an input is executable.
Candidate rejection represents an offline filtering mechanism and is fundamentally distinct from a runtime safety failure during test execution.

\textbf{Random search} draws candidates according to each representation's
sampling rule until its valid-test budget is filled. The geometry-aware
Manifold configuration samples an RCM-compatible direction and an insertion depth along
the forward ray inside the workspace, conditioned on robot feasibility.
For the other spaces, we uniformly sample points along each coordinate axis.

\textbf{Traversal} enumerates a discretized domain in each representation,
using $119\times119\times7$ grid counts to align the per-axis resolution.
Cartesian and PCA traversal enumerate rectangular coordinate grids and reject infeasible
points offline.
For all the three representations, a cell at the boundary of feasible and infeasible areas receives
up to 128 uniform sampling attempts within its coordinate bounds; the first
feasible input is retained for this cell.
This yields 5,075 Manifold reference cells.

\textbf{GA} search directs test generation using domain prior risk distributions and dynamic runtime failure feedback within each candidate space. The Manifold configuration searches
$(\theta_1,\theta_2,s)$, where the search is initialized using a subject-specific prior derived from earlier exploratory evaluations. Corresponding coordinate axes and prior distributions are similarly applied to the Cartesian XYZ and PCA representations. Because these test generation strategies rely on heuristic search, their behavior is inherently coupled with the underlying spatial representation. Changing the underlying spatial representation fundamentally alters the physical search landscape and the geometric effect of genetic operators, including mutation, crossover, and initial population sampling.

\section{Results and Analysis}

\subsection{RQ1: Candidate Validity and Generation Efficiency}
\label{sec:rq1-validity}

A valid case satisfies the input feasibility checks and can be submitted to
the simulator. We define candidate validity as the fraction of valid inputs among
candidates proposed during test generation, excluding domain construction and training-data generation,
formally, $N_{\mathrm{valid}}/N_{\mathrm{candidate}}$.
\autoref{tab:rq1-random-validity} reports the random-generation counts across
four runs per space. Every space ultimately executes $18412$ valid cases,
but as different spaces have different restrictions, the validity can vary.
Total time includes all the steps for each space to produce cases.
For a fair comparison, we measure the total execution time on the same hardware platform.
Apart from the random sampling time, Manifold counts the one-time walking of Manifold direction cells before sampling;
and PCA also counts time for generating training-input data and training the model.

\begin{table*}[htbp]
\centering
\caption{Random-generation candidate validity and total generation time aggregated over four runs.}
\label{tab:rq1-random-validity}
\begin{tabular}{lrrrrr}
\toprule
Space & Runs & Valid cases & Candidate cases & Candidate validity & Total time (s) \\
\midrule
Cartesian XYZ & 4 & 18,412 & 384,270 & $4.79\%$ & 12.43 \\
Manifold & 4 & 18,412 & 18,412 & $100.00\%$ & 2.15 \\
PCA & 4 & 18,412 & 75,117 & $24.51\%$ & 50.62 \\
\bottomrule
\end{tabular}
\end{table*}

Candidate tests generated from the feasible test domain achieve 100\% validity because 
robot kinematics and task-space constraints are intrinsically encoded during domain construction.
Sampling directly from this feasible test domain guarantees feasibility, whereas candidate proposals in Cartesian XYZ and PCA latent spaces frequently violate physical constraints and are rejected offline. Consequently, the Manifold representation yields significantly lower total test generation times compared to alternative spaces. Conversely, PCA exhibits the highest computational overhead due to its multi-stage pipeline: it requires sampling initial training configurations from Cartesian space, fitting the lower-dimensional latent model, and executing subsequent latent-space search.
Furthermore, because PCA sampling does not inherently preserve task-space constraints, it incurs substantial runtime penalties from iterative rejection sampling.

\noindent\textbf{Answer to RQ1.}
The constructed feasible test domain achieves 100\% candidate validity, outperforming Cartesian XYZ (4.79\%) and PCA (24.51\%). Eliminating candidate rejections allows a fixed sampling budget to produce significantly more executable tests.

\subsection{RQ2: Effectiveness of Test Space Representations across Testing Strategies}
\label{sec:rq2-representation}

\subsubsection{Budgets and Metrics}

\autoref{tab:rq2-performance} summarizes the total executed tests, reference-grid
coverage, and failure-pattern counts for the three  representations under each testing strategy.
Random and GA are evaluated over four runs of 4,603 valid test cases per
space to mitigate sampling randomness. Traversal has one run per space.

For traversal, we use the same per-axis grid counts ($119\times119\times7$) across the three input spaces.
This resolution alignment can result in different numbers of executable tests due to differences in the geometry of their feasible domains.

Reference-grid coverage is the fraction of reference cells containing at least one tested input.
For random and GA, coverage is reported as the average over four runs; for traversal,  it is the single-run value.

We classify failed trajectories using the recorded geometric states after withdrawal begins.
Each state sequence is truncated at the first safety violation, and consecutive repetitions of the same state are collapsed.
The resulting sequences define the seven observed patterns P1--P7 in \autoref{tab:rq2-performance}.
For Random and GA, the table sums the number of cases in each pattern across four runs.

\begin{table}[t]
\centering
\caption{Executed tests, reference-grid coverage, and failure-pattern counts by strategy and space.
Traversal uses resolution-aligned configurations.}
\label{tab:rq2-performance}
\footnotesize
\setlength{\tabcolsep}{3pt}
\begin{tabular}{@{}llrr@{\hspace{13pt}}*{7}{r}@{}}
\toprule
Strategy & Space & Executed tests & Coverage (\%) & \multicolumn{7}{c}{Failure patterns (count)} \\
\cmidrule(lr){5-11}
 & & & & P1 & P2 & P3 & P4 & P5 & P6 & P7 \\
\midrule
Random & Cartesian XYZ & $4\times4{,}603$ & 51.03 & 943 & 0 & 0 & 0 & 0 & 0 & 0 \\
 & Manifold & $4\times4{,}603$ & 52.22 & 609 & 5 & 3 & 2 & 4 & 1 & 0 \\
 & PCA & $4\times4{,}603$ & 49.96 & 409 & 0 & 0 & 0 & 0 & 0 & 0 \\
\midrule
Traversal & Cartesian XYZ & $1\times3{,}846$ & 47.78 & 382 & 1 & 1 & 0 & 0 & 0 & 0 \\
 & Manifold & $1\times5{,}075$ & 100.00 & 455 & 0 & 1 & 2 & 0 & 0 & 0 \\
 & PCA & $1\times30{,}027$ & 55.61 & 999 & 1 & 0 & 0 & 0 & 0 & 0 \\
\midrule
GA & Cartesian XYZ & $4\times4{,}603$ & 21.33 & 12,521 & 5 & 0 & 0 & 0 & 0 & 0 \\
 & Manifold & $4\times4{,}603$ & 11.72 & 16,027 & 392 & 8 & 31 & 13 & 5 & 1 \\
 & PCA & $4\times4{,}603$ & 9.34 & 14,406 & 2 & 0 & 0 & 0 & 0 & 0 \\
\bottomrule
\end{tabular}
\par\smallskip
\parbox{\linewidth}{\footnotesize
Executed tests are shown as runs $\times$ tests per run.
Random and GA coverage is averaged over four runs; traversal reports a single run.
All coverage values use the same 5,075-cell reference grid.
P1--P7 are mutually exclusive first-violation patterns.
Geometric states: $I$: inside the eye; $R$: permitted entry channel; $O$: outside the eye; $F$: forbidden region.
Arrows denote transitions between consecutive distinct recorded states.
\par\smallskip
\begin{tabular}{@{}p{0.05\linewidth}p{0.24\linewidth}p{0.66\linewidth}@{}}
P1 & $I\rightarrow F$ & Enters the forbidden region from inside the eye, without first visiting the entry channel or the outside. \\
P2 & $F$ & The first recorded point is already in the forbidden region; no safe motion prefix is observed. \\
P3 & $I\rightarrow O$ & Crosses from inside to outside the eye; the straight segment joining the two recorded points does not safely pass through the permitted entry channel. \\
P4 & $I\rightarrow R\rightarrow F$ & Enters the permitted entry channel from inside the eye, then enters the forbidden region. \\
P5 & $I\rightarrow R\rightarrow O\rightarrow F$ & Exits the eye through the entry channel, then enters the forbidden region from outside. \\
P6 & $I\rightarrow R\rightarrow O\rightarrow R\rightarrow F$ & Exits the eye, returns to the permitted entry channel, then enters the forbidden region. \\
P7 & $R\rightarrow F$ & The first recorded point is in the permitted entry channel; the tip subsequently enters the forbidden region. \\
\end{tabular}}
\end{table}

\subsubsection{Comparisons}

We compare spatial coverage and failure discovery using the common reference grid
and the failure patterns in \autoref{tab:rq2-performance}. The reference grid
contains 5,075 cells derived from the Manifold discretization. Manifold traversal
therefore achieves $100\%$ reference-grid coverage by construction.

\paragraph{Random}
Under the same execution budget, Cartesian XYZ yields more failures than Manifold, although their reference-grid coverage difference is small.
However, all 943 Cartesian failures follow P1, whereas the 624 Manifold failures include six patterns.
PCA likewise reveals only P1, with fewer failures.
Thus, similar cell coverage can conceal differences in the failure behaviors
exercised, and a larger failure count can reflect repeated observations of
one behavior rather than a wider range of unsafe executions. The additional rare Manifold patterns show that successful passage through the channel does not
ensure safety over the remainder of the trajectory. The evidence shows that our method has a broader behavioral diversity.

\paragraph{Traversal}
Under aligned per-axis grid counts, PCA executes substantially more tests
than Manifold yet covers fewer reference cells, indicating that additional
executions do not necessarily broaden spatial exploration.

Manifold and Cartesian traversal both reveal three patterns, but their
discoveries differ: only Manifold reveals failure after entering the channel (P4),
while Cartesian additionally observes P2.
However, complete reference-grid coverage of Manifold traversal still misses P2 and P5--P7. Different inputs within the same cell may exhibit different failure behaviors, so the coverage alone cannot establish behavioral completeness. This highlights the importance of considering coverage together with failure yield and observed behavioral diversity when assessing test results.

\paragraph{GA}
Manifold GA combines the highest failure yield with broader observed
pattern diversity, despite covering fewer reference cells than Cartesian
GA. Manifold GA discovers additional P4--P7 cases, extending the behavioral
evidence to channel- and exit-related violations absent from both baselines.

GA primarily increases observations of dominant failure behaviors
while also reaching rare ones. It is designed to concentrate on failure-prone regions with the
use of risk priors and runtime failure feedback.

\subsubsection{Pass and Fail Outcome Distributions}

\begin{figure}[t]
\centering
\begingroup
\definecolor{rqTwoPass}{HTML}{3277A5}
\definecolor{rqTwoFail}{HTML}{D95327}
\normalsize
\tikz[baseline=-0.5ex]{\fill[rqTwoPass] (0,0) circle (2.5pt);}\, Pass
\hspace{2em}
\tikz[baseline=-0.5ex]{\draw[rqTwoFail,line width=1pt] (-3pt,-3pt)--(3pt,3pt) (-3pt,3pt)--(3pt,-3pt);}\, Fail
\par
\endgroup
\vspace{0.2em}
\begin{minipage}{0.325\linewidth}
  \centering
  \includegraphics[width=\linewidth]{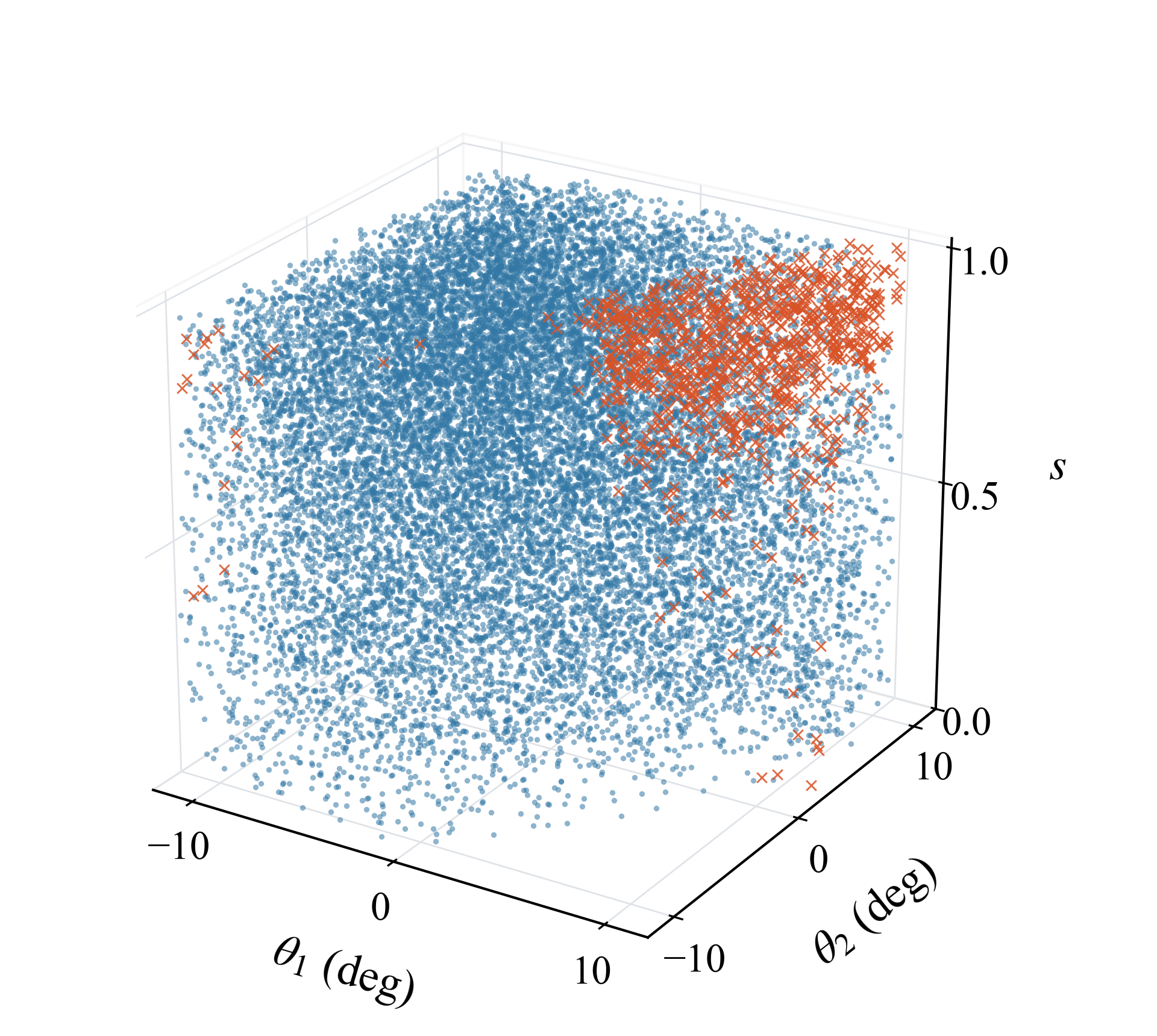}\\[-0.25em]
  \footnotesize (a) Cartesian XYZ: Random
\end{minipage}\hfill
\begin{minipage}{0.325\linewidth}
  \centering
  \includegraphics[width=\linewidth]{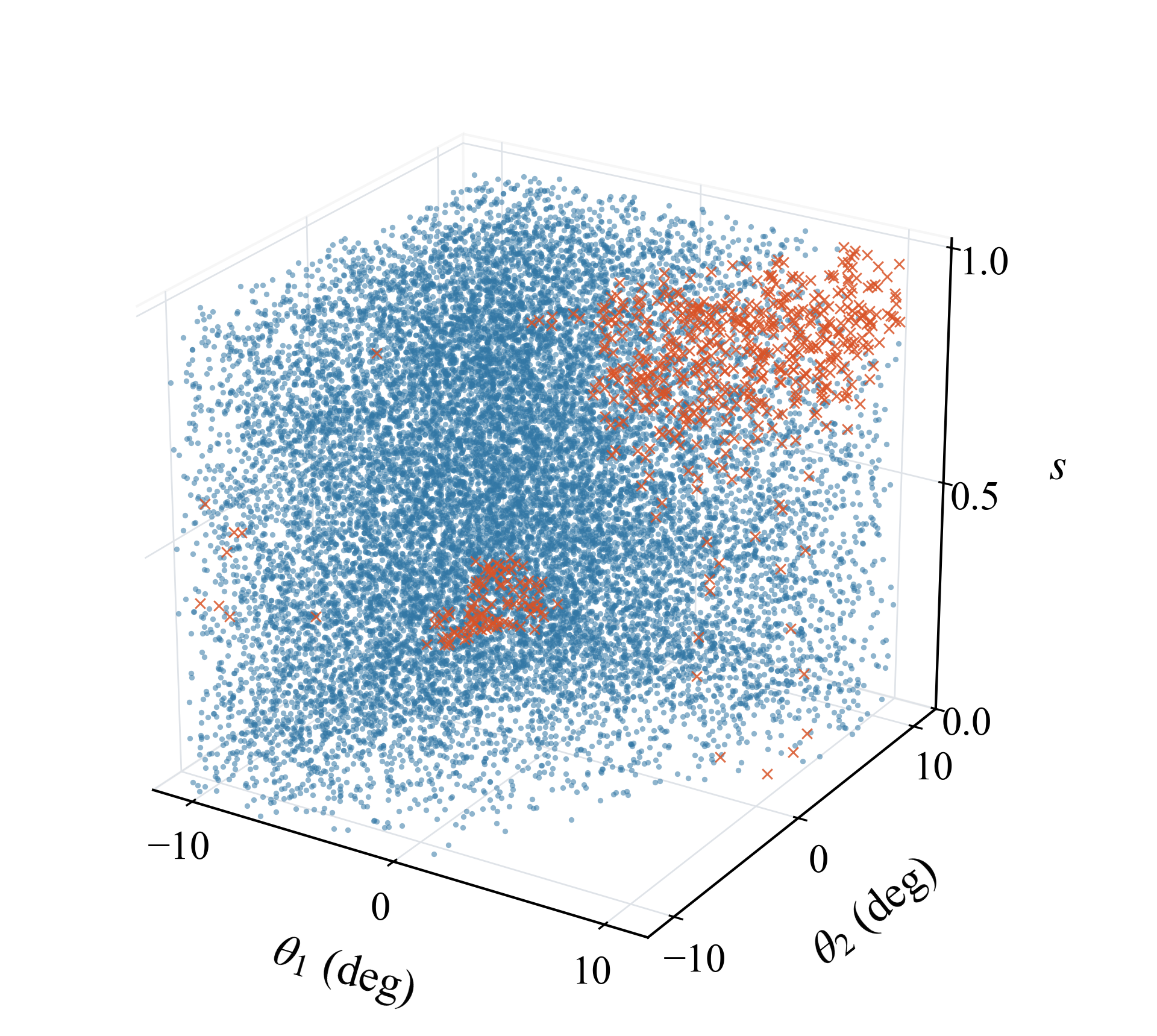}\\[-0.25em]
  \footnotesize (b) Manifold: Random
\end{minipage}\hfill
\begin{minipage}{0.325\linewidth}
  \centering
  \includegraphics[width=\linewidth]{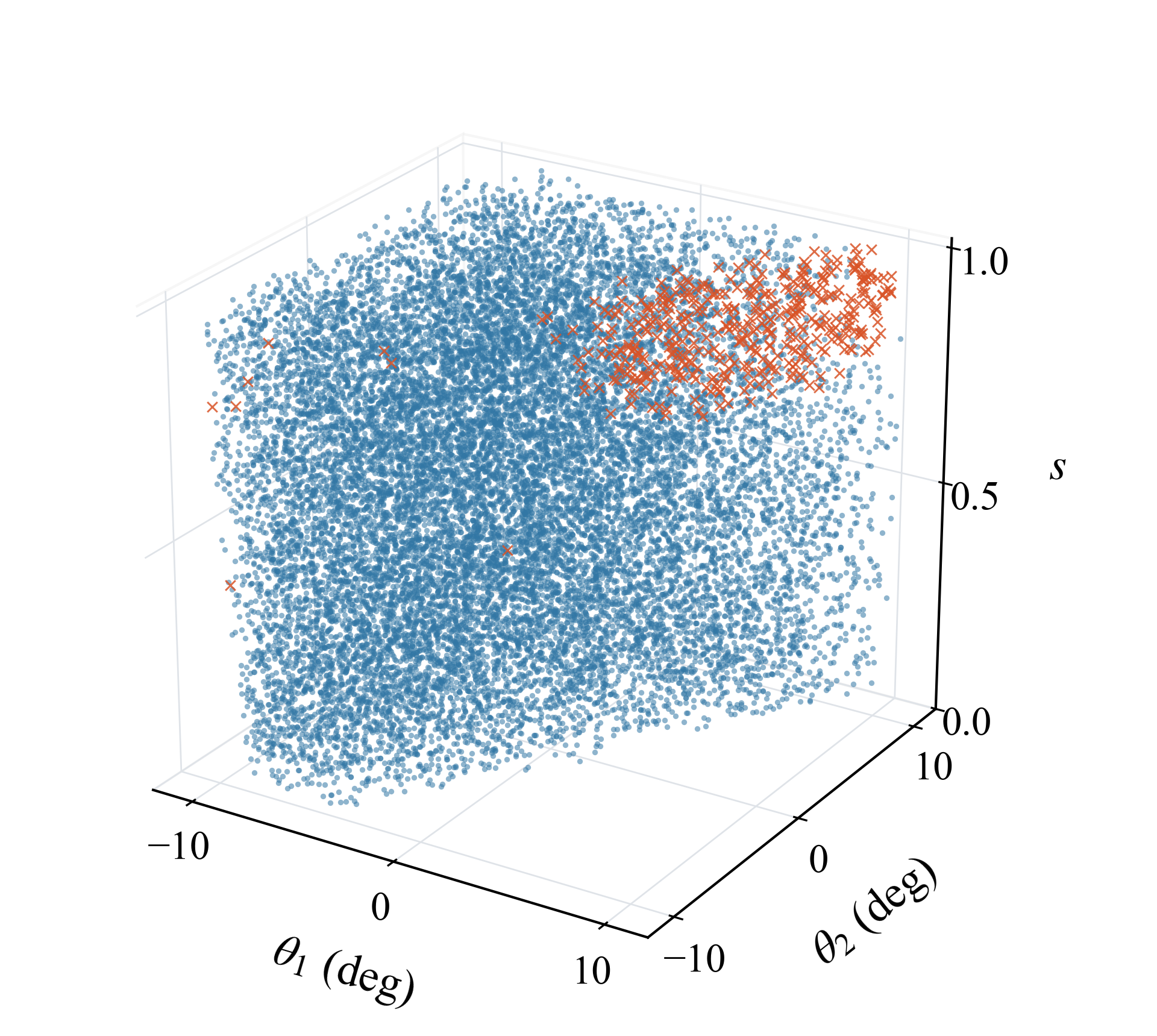}\\[-0.25em]
  \footnotesize (c) PCA: Random
\end{minipage}

\vspace{0.6em}
\begin{minipage}{0.325\linewidth}
  \centering
  \includegraphics[width=\linewidth]{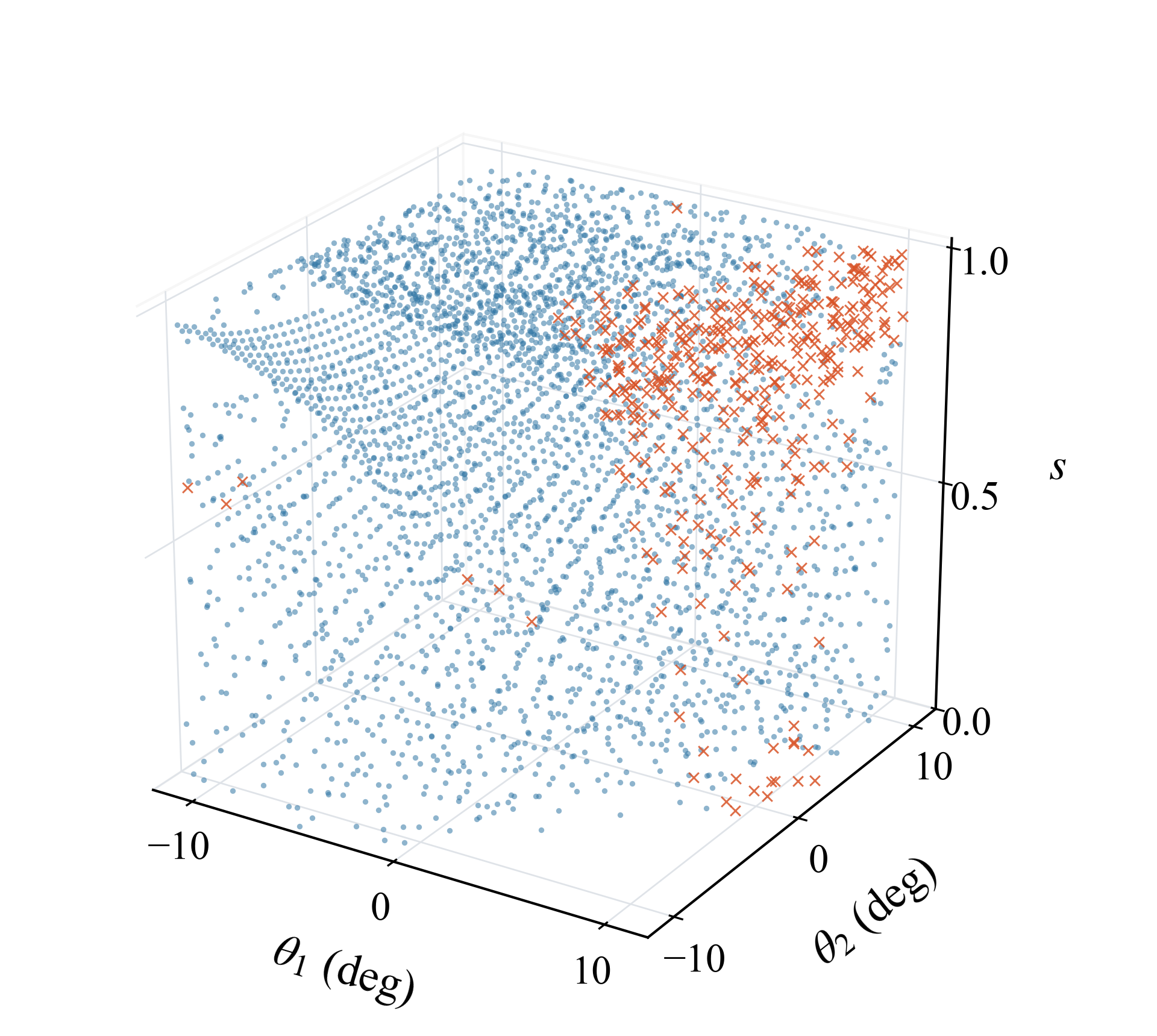}\\[-0.25em]
  \footnotesize (d) Cartesian XYZ: Traversal
\end{minipage}\hfill
\begin{minipage}{0.325\linewidth}
  \centering
  \includegraphics[width=\linewidth]{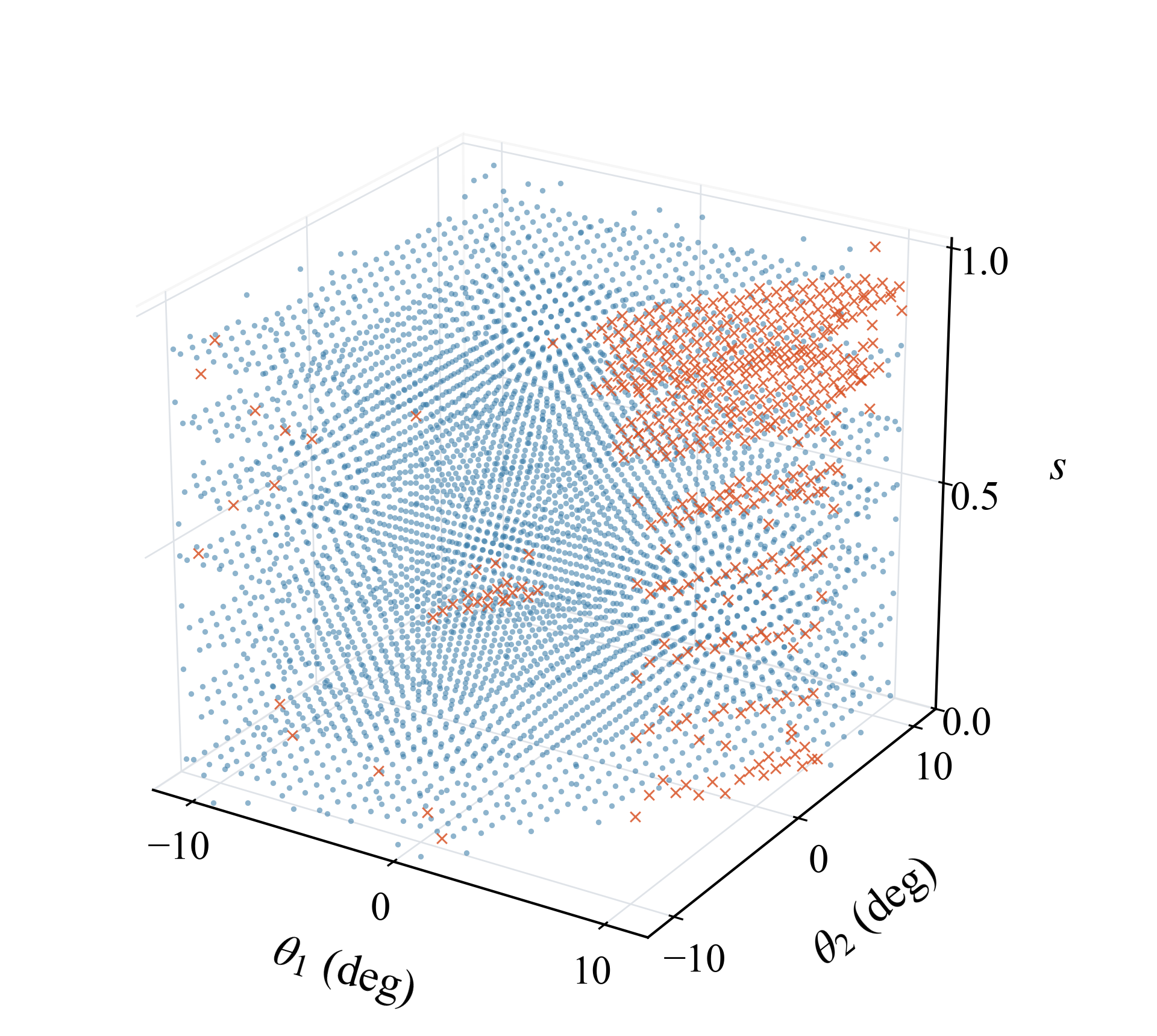}\\[-0.25em]
  \footnotesize (e) Manifold: Traversal
\end{minipage}\hfill
\begin{minipage}{0.325\linewidth}
  \centering
  \includegraphics[width=\linewidth]{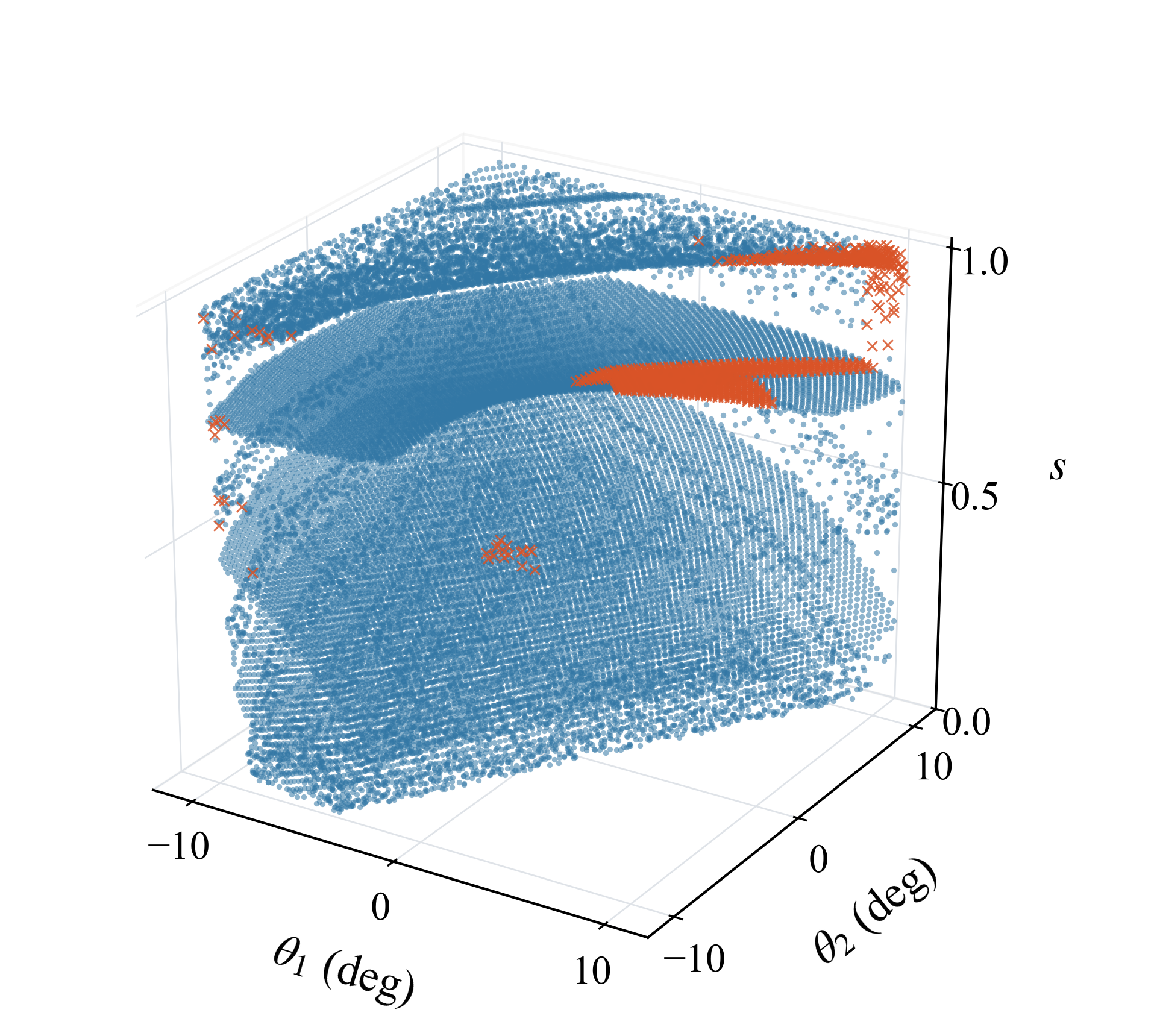}\\[-0.25em]
  \footnotesize (f) PCA: Traversal
\end{minipage}

\vspace{0.6em}
\begin{minipage}{0.325\linewidth}
  \centering
  \includegraphics[width=\linewidth]{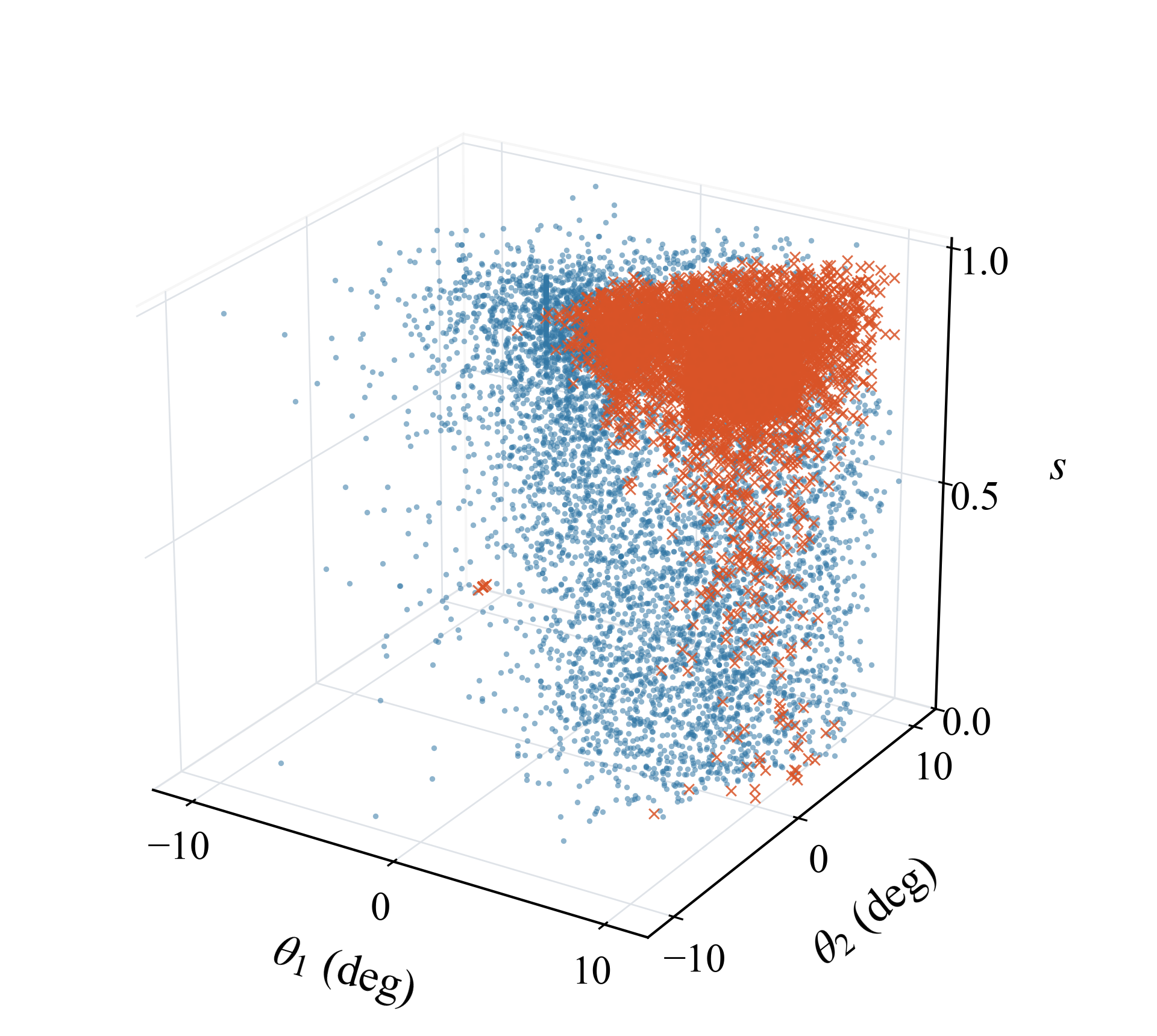}\\[-0.25em]
  \footnotesize (g) Cartesian XYZ: GA
\end{minipage}\hfill
\begin{minipage}{0.325\linewidth}
  \centering
  \includegraphics[width=\linewidth]{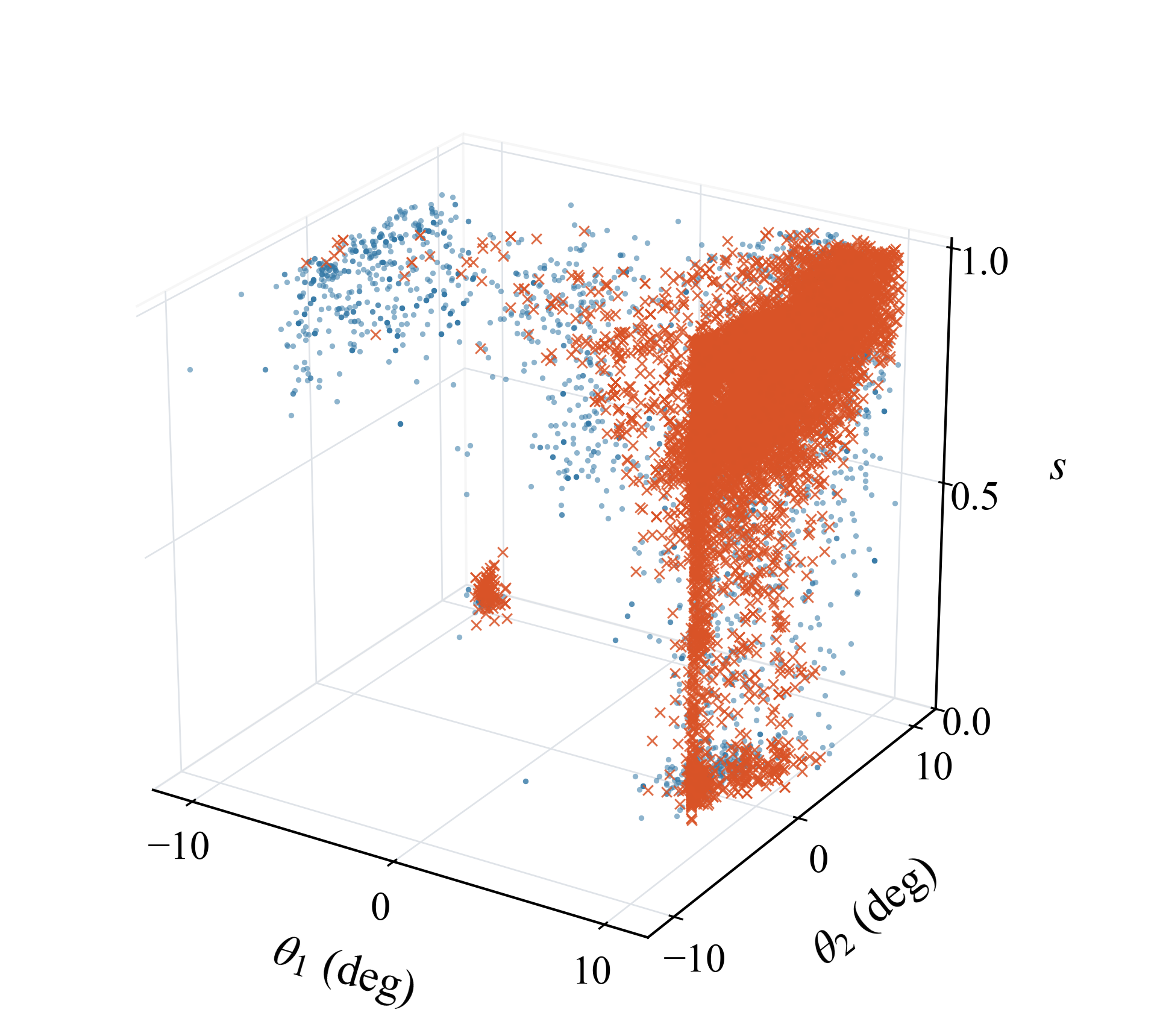}\\[-0.25em]
  \footnotesize (h) Manifold: GA
\end{minipage}\hfill
\begin{minipage}{0.325\linewidth}
  \centering
  \includegraphics[width=\linewidth]{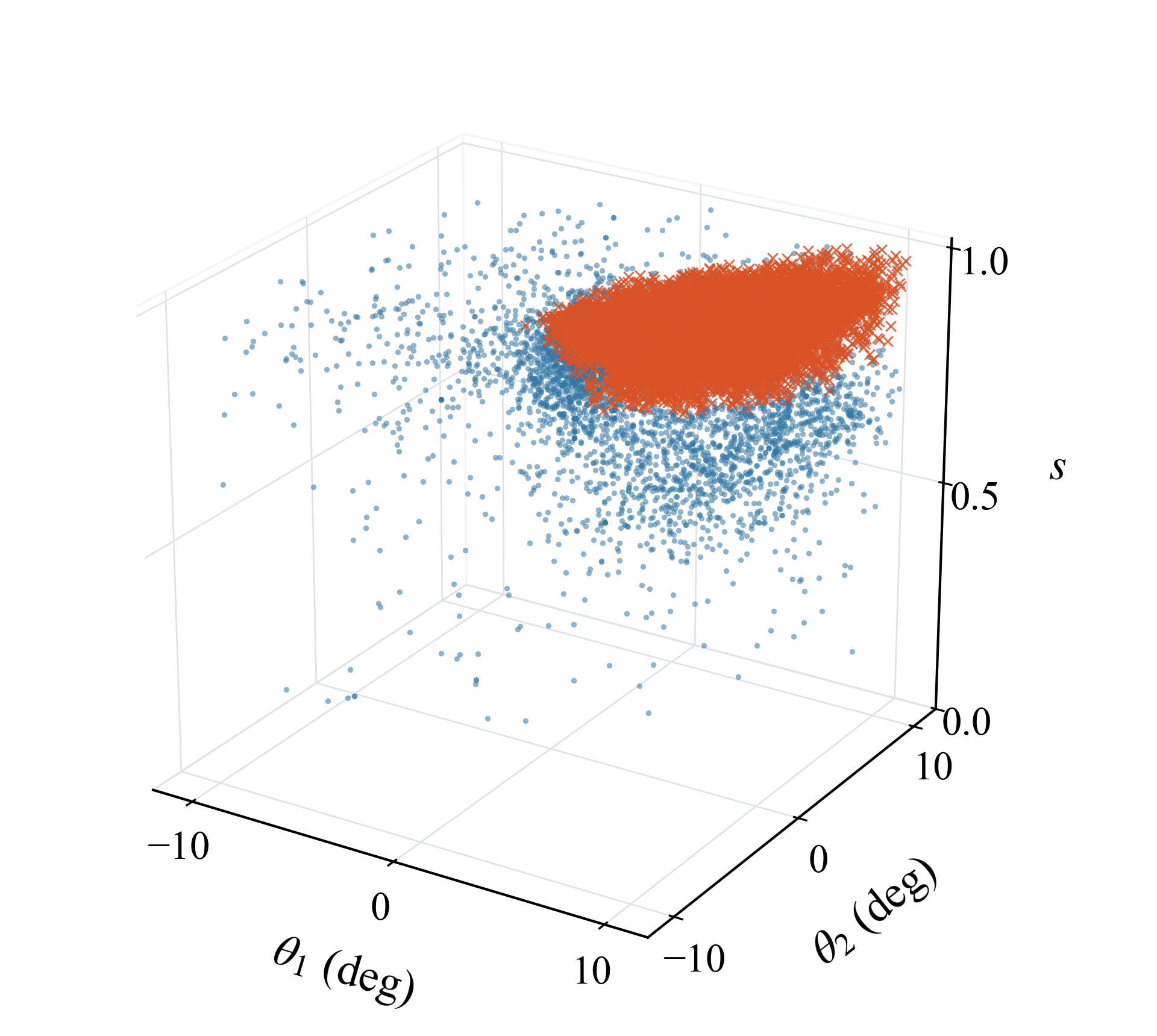}\\[-0.25em]
  \footnotesize (i) PCA: GA
\end{minipage}
\caption{Pass and fail distributions across three strategies and three spaces in common task coordinates.}
\label{fig:rq2-outcome-distributions}
\Description[Nine pass and fail distributions by strategy and space]{%
A three-by-three matrix of three-dimensional scatter plots with shared
axes for theta-one, theta-two, and normalized insertion depth. Rows show
random search, traversal, and GA; columns show Cartesian XYZ, Manifold,
and PCA. Blue dots indicate passes and orange crosses indicate failures.
Random testing spreads samples broadly, traversal reveals space-dependent
sampling patterns, and GA concentrates samples in failure-prone regions.}
\end{figure}

\autoref{fig:rq2-outcome-distributions} shows pass and fail distributions
for all nine strategy--space combinations in the common task coordinates
$(\theta_1,\theta_2,s)$. Rows correspond to random search, traversal, and GA;
columns correspond to Cartesian XYZ, Manifold, and PCA. All panels use the same
axis limits and viewing angles and preserve each method's sampling pattern.
These panels illustrate the inputs that produce passing and failing executions.

The random-search row shows broad sampling with localized failure regions.
Manifold samples cover two major failure regions and additional failures
near both angular boundaries of $\theta_1$, whereas Cartesian and PCA
samples cover only one major failure region.

Traversal makes the effect of the input representation particularly visible.
All three spaces reach the main failure region. However, their sampling
patterns differ in the common task coordinates: Manifold traversal follows
the angular and normalized-depth grid, while the Cartesian and PCA grids
map to unevenly spaced layers. The dense PCA layers illustrate why a much
larger test set can still cover only part of the reference grid.

GA search concentrates test evaluations within failure-prone regions, producing distinct spatial clustering patterns across the three candidate spaces. All three representations successfully locate the primary error-prone region. However, only the Cartesian XYZ and Manifold spaces effectively expose a secondary failure region near the lower boundary of $\theta_1$.

\noindent\textbf{Answer to RQ2.}
The Manifold representation supports systematic exploration of the feasible
task domain and reveals a broader range of observed failure behaviors under
random and GA testing. Manifold random testing discovers six patterns
despite producing fewer failures than Cartesian random testing, and
Manifold GA discovers all seven with the highest failure rate. The channel-
and exit-related patterns P4--P7 are observed only with Manifold in these
experiments. Traversal visits every Manifold-derived reference cell by
construction, although its failure rate is lower than Cartesian traversal
and both discover only three patterns. Together, these results show that
task-space representation affects both the testing strategies and the
failure behaviors they expose.

\subsection{RQ3: Real Defect Detection and Transferability}
\label{sec:rq3-effectiveness-transfer}

RQ3 evaluates whether the proposed testing framework provides actionable empirical evidence to expose a realistic controller defect, and assesses its transferability when instantiated on an alternative surgical robotic platform.

\subsubsection{Statistical Detection of a Real Controller Defect}

To evaluate whether the workflow exposes a defect in executable controller software, we compare a defective implementation (v1) against a patched version (v2). Controller v1 contains a rare defect that triggers only when two motors approach their physical limits.  We evaluate each controller with random search and
GA, using four runs of $4603$ simulations for every controller--strategy pair.
Table~\ref{tab:rq3-controller-comparison} reports run-level means and standard
deviations under this common rule.

\begin{table*}[t]
\centering
\caption{Results from controllers v1 and v2.  ``Motor-limit region'' means that two of the motors lie within the outer $10\%$ of their valid motor ranges.}
\label{tab:rq3-controller-comparison}
\footnotesize
\setlength{\tabcolsep}{4pt}
\begin{tabular}{@{}llrrrr@{}}
\toprule
Controller & Strategy & Runs $\times$ budget & Failures &
Mean failure rate $\pm$ SD & Failures in motor-limit region \\
\midrule
v2 & Random & $4\times4603$ & 624   & $3.39\%\pm0.08$ pp & $32/753$ ($4.25\%$) \\
v1 & Random & $4\times4603$ & 613   & $3.33\%\pm0.33$ pp & $47/775$ ($6.06\%$) \\
v2 & GA     & $4\times4603$ & 16477 & $89.49\%\pm0.49$ pp & $9432/10062$ ($93.74\%$) \\
v1 & GA     & $4\times4603$ & 16803 & $91.26\%\pm0.67$ pp & $10219/10698$ ($95.52\%$) \\
\bottomrule
\end{tabular}
\end{table*}

Although the defect is known from the original controller implementation, individual oracle violations cannot be definitively attributed to this specific fault, as other controller faults can yield identical failure outcomes. 
We therefore compare failure statistics between v1 and v2 to assess whether
testing reveals a difference associated with the repair. Random testing
yields nearly identical overall failure rates across the two versions.
GA concentrates tests in failure-prone regions and reveals a mean failure
rate approximately 1.8 percentage points higher for v1 across four runs.

The motor-limit results help interpret this difference. With the same number of simulated cases,
GA provides roughly 13--14 times as many observations in the region
associated with the known defect. Within this region, the v1
controller has a higher failure rate under both strategies. GA's contribution
is the much larger number of relevant tests and a clearer difference in
overall failure rates. The observed pattern is consistent with the known
defect, although the aggregate statistics cannot identify which individual
failures it caused.

This case study shows how the manifold-based workflow can investigate a
real controller defect under the interference of other failures. The feasible test domain supplies
executable inputs, and the mapping to motor commands identifies tests near
the relevant motor limits. Random search broadly samples this domain,
while GA concentrates tests in failure-prone configurations. Comparing the
two controllers across these strategies yields statistical evidence of a
change associated with the repair, which can also support regression testing.

\subsubsection{Transfer to a Different Robot With the Same Task}
\label{sec:rq3-transfer}

The transfer experiment investigates whether decoupling task space geometry from robot-specific kinematics enables workflow portability beyond the five-motor mechanism. To evaluate transferability, we instantiate the framework on a secondary robotic platform: a 7-degree-of-freedom Kinova Gen3 manipulator equipped with a Robotiq gripper, integrated via MoveIt 2 and ROS 2 Jazzy. The new platform performs the same task specified in \autoref{sec:example}. This case study assesses workflow generalizability to kinematically redundant serial manipulators, where multiple joint configurations can satisfy a given end-effector target pose. Because this evaluation serves primarily as a proof-of-concept for workflow transferability, we deliberately omit exhaustive comparative evaluations across all Cartesian XYZ and PCA baseline configurations conducted in RQ2.

\paragraph{Transfer Design}
The transfer preserves the three task coordinates $(\theta_1,\theta_2,t)$
and constructs the feasible test domain for the new robot.
The five-motor analytic IK of our previous robot is replaced by a Kinova-specific seeded numerical IK.
Because seven-joint IK is not unique, we fix the IK branch selection policy to always
choose the solution with the smallest joint-space movement from the current state.

We implement an adapter to submit test inputs to MoveIt~2 and retrieve
planned trajectories for offline evaluation. The formal evaluation is
\emph{plan-only}: OMPL (the motion planning library integrated within MoveIt~2) plans the withdrawal trajectory, and forward
kinematics maps the returned joint trajectory to tip positions for the
same geometric safety checks. We encode the RCM entry constraint in the
planner's target model through forbidden regions.
\paragraph{Constructed domain and completed tests.}

As the IK is not analytically solved for the new robot, there exists a numerically sensitive boundary.
We use a $150\times150$ angular grid with $31$ depth probes per direction to identify
a set of $9762$ feasible insertion-depth intervals after excluding the numerically sensitive
collision boundary. The reference grid subdivides each retained insertion-depth interval into
7 depth bins, yielding $68334$ reference cells.

We conducted two testing strategies on this system.
Table~\ref{tab:kinova-strategy-results} summarizes the completed runs.
We count the total number of test inputs sent to the planner. Failures comprise planning
failures and task failures on returned trajectories.
Planning failure means that the tested planner reports failure to produce a valid trajectory.
Task failure means that a returned trajectory violates the task constraints checked by the oracle.
Failure rate is total failures divided by tests.
Reference-grid coverage is the fraction of the $68334$ reference cells assigned at least one tested input, including inputs resulting in failures.

\begin{table}[t]
  \centering
  \caption{Results of Kinova plan-only runs.}
  \label{tab:kinova-strategy-results}
  \small
  \setlength{\tabcolsep}{3pt}
  \begin{tabular}{@{}lrrrrrr@{}}
    \toprule
    Strategy & Tests & \multicolumn{3}{c}{Failures} & Failure rate & Coverage \\
    \cmidrule(lr){3-5}
     & & Planning & Task & Total & & \\
    \midrule
    Random & 68,918 & 8,204 & 3,615 & 11,819 & $17.15\%$ & $63.78\%$ \\
    Traversal & 68,334 & 8,229 & 3,067 & 11,296 & $16.53\%$ & $100.00\%$ \\
    \bottomrule
  \end{tabular}
\end{table}

The two strategies achieve similar failure rates but differ substantially in
coverage. Traversal assigns a test to every reference cell, whereas random
testing revisits some cells and leaves others unexplored despite a comparable
test budget. Random testing records slightly more failures, primarily from
task violations rather than planning failures.
This is because random distributes test inputs differently across the feasible domain,
whereas traversal testing is restricted to points on a fixed-resolution grid.
As a result, random testing may sample more inputs from regions where task constraints are violated.
Both strategies exercise the
transferred domain and record planning and task failures in a common
reference grid; this single-run comparison illustrates the transferability of
our workflow between different robots with the same task.

\noindent\textbf{Answer to RQ3.}
The case studies demonstrate the practical utility of manifold-based testing
for exposing a real controller defect and transferring the workflow between
robots performing the same task. On the original robot, GA searches the
feasible test domain and concentrates tests near the motor limits that
trigger the known defect, revealing a difference between two versions of the
controllers that is obscured in the overall random-testing results.
On the Kinova Gen3, the workflow keeps the task coordinates while adapting
robot-specific realization to numerical IK and MoveIt~2 planning.
Both random testing and traversal generate tests and reveal planning and
task failures on this robot. Together, these results support
using task coordinates as a reusable basis for test generation and failure
analysis across different kinematic structures, with robot-specific
realizability checks and result evaluation.

\section{Discussion}

\subsection{Feasibility and Testing Efficiency}

Feasibility-aware generation reduces rejected candidates, but its testing
benefit depends on the budget constraint. For a fixed number of valid tests (RQ1),
higher validity requires fewer candidates; under a fixed execution budget (RQ2),
coverage and failure discovery depend on testing strategy rather than validity alone.
Overall efficiency should also account for domain construction costs, such as PCA offline preprocessing. Consequently, feasibility-aware pruning is a necessary prerequisite—but not a sufficient guarantee—for effective and efficient test generation.

\subsection{Complementary Testing Strategies}

Failure yield, spatial coverage, and behavioral diversity capture different
aspects of test effectiveness and should be assessed together. Traversal
provides a reference-cell baseline, random testing explores broadly within cells,
and GA concentrates on anticipated or observed hazards.
The additional patterns found by Manifold random and GA testing in RQ2 show
that visiting every reference cell does not exhaust its possible behaviors.
These findings suggest combining coverage metrics with both exploration and targeted search.
The controller comparison (\autoref{tab:rq3-controller-comparison}) further illustrates how targeted
testing can reveal aggregate differences associated with a repair.

\subsection{Threats to Validity}

\textbf{Internal validity.} Within the  five-motor evaluation, all experimental configurations share
simulator and safety-oracle implementations and input-feasibility criteria.
However, equivalent sampling or mutation rules in different coordinate
spaces cannot ensure equivalent distributions of physical robot states.
The comparisons therefore reflect the  combined effect of the chosen input space and strategies.
Furthermore, our approach relies on constructing an appropriate task manifold and defining its feasible test domain, which can demand substantial domain-modeling effort. Identifying a task-space representation that effectively encapsulates failure-relevant variables while remaining computationally tractable can be challenging. Even when suitable task coordinates are available, complex workspace geometries and forbidden-region constraints may preclude an exact closed-form characterization of the feasible domain, necessitating numerical approximations and runtime feasibility checks. Omissions or errors in domain construction risk either excluding critical physical states or admitting infeasible inputs, thereby distorting both failure discovery metrics and coverage estimates. The observed performance benefits thus depend on the fidelity of the constructed domain, and offline construction costs may offset net efficiency gains.

 \textbf{External validity.} 
Our primary evaluation considers a simulated five-motor robot, while the Kinova case study demonstrates cross-platform transferability to a kinematically redundant manipulator within an open-loop, motion-planning framework. Extension to other surgical tasks or physical execution remains unverified and requires suitable task coordinates, realization checks, and trajectory oracles.
For testing on physical robots, the feasible task domain could guide the selection of executable configurations, while different strategies could help prioritize tests under limited hardware execution budgets. In addition to the offline trajectory oracle, such deployment would require physical feasibility checks, reliable robot initialization, and online safety monitoring with protective stopping mechanisms. Whether the efficiency and failure-discovery benefits observed in simulation carry over to physical execution remains to be evaluated.
The workflow could also be adapted to evaluate Vision-Language-Action (VLA) models such as OpenVLA~\cite{kim2025openvla}, where feasible task states initialize scenes, rendered observations and instructions feed the policy, and generated trajectories are monitored by the oracle. Because geometry alone cannot capture variations in visual rendering, language phrasing, or interaction history, evaluating VLA policies requires additional test dimensions and multimodal coverage metrics.

\section{Conclusion}

This paper presents a manifold-centered testing framework that separates feasible task-domain construction from testing strategies. On an RCM-constrained robotic task, the evaluation shows that task coordinates support efficient generation of executable inputs, while complementary testing strategies provide evidence of spatial coverage and diverse failure behaviors. The controller comparison and Kinova case study further demonstrate the framework's utility for investigating a known defect and transferring the workflow across robot kinematics within the same task. These findings support feasibility-aware task modeling as a basis for systematic robot testing; applicability to broader tasks and physical execution remains to be evaluated.

\bibliographystyle{ACM-Reference-Format}
\bibliography{references}

\end{document}